\documentclass[twocolumn,prc,preprintnumbers,nofootinbib,superscriptaddress,showpacs,a4]{revtex4-1}
\usepackage{epsfig,graphicx,float,pst-all}
\usepackage{mathrsfs}
\usepackage[utf8]{inputenc}
\usepackage{rotating}
\usepackage{url}
\usepackage{esvect}
\usepackage{subfigure}
\usepackage{multirow}
\usepackage{amsmath}
\usepackage{amssymb}
\usepackage{color}

\usepackage{natbib}
\usepackage{hyperref}
\usepackage[warn]{textcomp}
\usepackage{gensymb}
\usepackage{footnote}
\usepackage{siunitx}
\usepackage{comment}
\usepackage{bm}
\usepackage[normalem]{ulem}
\begin{document}

\preprint{NITEP 71}
\title{Systematic study on the role of various higher-order processes in the breakup of weakly-bound projectiles}
\author{Jagjit Singh}
\email{jsingh@rcnp.osaka-u.ac.jp}
\affiliation{Research Center for Nuclear Physics (RCNP), Osaka University, Ibaraki 567-0047, Japan}
\author{Takuma Matsumoto}
\email{matsumoto@phys.kyushu-u.ac.jp}
\affiliation{Department of Physics, Kyushu University, Fukuoka 819-0395, Japan}
\author{Kazuyuki Ogata}
\email{kazuyuki@rcnp.osaka-u.ac.jp}
\affiliation{Research Center for Nuclear Physics (RCNP), Osaka University, Ibaraki 567-0047, Japan}
\affiliation{Department of Physics, Osaka City University, Osaka 558-8585, Japan}
\affiliation{Nambu Yoichiro Institute of Theoretical and Experimental Physics (NITEP), Osaka City University,
Osaka 558-8585, Japan}
\date{\today}

\begin{abstract}
The  virtual photon theory (VPT), which is based on first-order Coulomb dissociation restricted to the electric dipole ($E1$), 
has been successfully used to explain the breakup data for several cases. 
Our aim is to study the role of various higher-order processes that are ignored in the VPT, 
such as the nuclear breakup, interference
between nuclear and Coulomb amplitudes, and multistep breakup processes mainly due
to strong continuum-continuum couplings in the breakup of two-body projectiles on a 
heavy target at both intermediate and higher incident energies. 
For the purpose of numerical calculations, we employed eikonal version of 
three-body continuum-discretized coupled-channels (CDCC) reaction model. 
Our results for the breakup of $^{11}$Be and $^{17}$F on $^{208}$Pb target at $100$, $250$, 
and $520$\,MeV/A, show the importance of nuclear breakup contribution, and its significant role in the multistep processes. 
The multistep effect on Coulomb breakup for core-neutron projectile was found to be negligible, 
whereas it was important for core-proton projectile. Coulomb-nuclear interference (CNI) effect was also found to be non-negligible.
Qantitatively, the multistep effects due to the nuclear breakup was found to depend on the incident energy through the energy dependence of the core-target and nucleon-target nuclear potentials.
The nuclear breakup component, the CNI effect, and the multistep breakup processes are all found to be non-negligible; hence, the assumptions adopted in the VPT for 
the accurate description of breakup cross sections are not valid. 
\end{abstract}
\maketitle
\section{Introduction}\label{INTRO}
Nuclear reactions are the main source of our present day understanding 
of atomic nuclei. The reactions with the unstable nuclei, lying away from the strip 
of stability on the eastern (neutron-rich) and western (proton-rich) sides 
of the nuclear chart, have opened up a new epitome in nuclear physics. 
This is due to the tremendous advancements in the radioactive 
ion beam facilities around the world \cite{Bennet00}.
At these facilities, high intensity beams of many unstable nuclei, which have 
very short half-lives and small one- or two-nucleon separation energies of order 
$1$\,-\,$2$ MeV, are produced and dedicated for nuclear reactions. 

Theoretically, this further stimulates the opportunity to improve on the existing 
understandings of both nuclear structure and reactions with these unstable nuclei, particularly 
in the proximity of driplines (limit of neutron or proton binding). 
These dripline systems flaunt the striking exotic phenomena, such as the formation of halo \cite{Tanihata85,Tanh85}, 
evolution of the new magic numbers \cite{Otsuka01}, and a narrow momentum distribution \cite{Kobayashi88}.
These weakly-bound nuclei that exhibit strong cluster-like structures can be described as a fragile system 
of a core plus one or two nucleons. Such type of unusual structures 
have earlier been observed in lighter nuclei, such as in He \cite{Tanh85}, Li \cite{Tanihata85}, 
Be \cite{Fukuda91,Palit03,Fukuda04}, B \cite{Warner95,Negoita96,Bazin98,Gulmaraeas00}, and C \cite{Bazin98,TOG16} isotopes as well as in 
relatively heavier nuclei, such as in Ne \cite{Nakamura2009}, Na \cite{Gaudefroy2012}, and Mg \cite{Kobayashi2014} isotopes. 
From the astrophysical perspective, these systems are of paramount importance and their properties serve as important inputs 
to the theoretical calculations on stellar burning, which otherwise are often forced to rely on global assumptions
about nuclear properties extracted from stable nuclei \cite{Bertulani10,Langanke13,Bertulani16}.

In most cases, these exotic nuclei have only one or two weakly-bound states, 
which make the couplings to the continuum significant.
Thus, conventional nuclear structure methods established
for stable nuclei cannot be directly extended to these unstable nuclei. Since these weakly-bound systems 
can be easily broken up in the nuclear and Coulomb fields of the heavy target, 
breakup reactions could serve as an effective tool to investigate the structure of these nuclei \cite{Yahiro12,Chatterjee18,Bonaccorso18}.
The  virtual photon theory (VPT), a model that assumes the first-order Coulomb dissociation 
with including only the electric dipole ($E1$) contribution, has been successful in 
explaining breakup observables for many cases \cite{Bertulani88,Baur96,Baur03}. 
Although the dissociation of the weakly-bound or a halo projectile is notably governed by the Coulomb interaction, 
the nuclear interaction with the target cannot be ignored in some cases \cite{Tanihata95}. 
The importance of the Coulomb-nuclear interference (CNI) effect, the consequences of the elimination of 
projectile-target nuclear interaction in the VPT and importance of multistep effects were discussed in 
Refs.~\cite{Hus06,Thompson09,Chatterjee02,Chatterjee03,Chatterjee07,Tarutina04,Mukeru15,Dasso98,Nunes98,Nunes99}. 
Also. there are a couple of comparative studies, which have been 
performed for the breakup of core-neutron ($\rm C$-$n$) versus core-proton ($\rm C$-$p$) projectiles 
on different targets and different beam energies \cite{Rangel16,Kumar11,Kucuk12,Paes12,Mukeru18}.
It is worthy to note that most of these studies in literature, discussed the importance of the 
higher-order effects at lower incident energies \textit{i.e.}, $\le 100$\,MeV/A. 
The relativistic effects on breakup are discussed in Ref.~\cite{Ogata09,Ogata10,Laura19}.
The appreciable modification in the reaction dynamics of a breakup of weakly-bound projectile is governed by the strong continuum-continuum and 
continuum-bound-state couplings. Higher-order couplings, which also VPT does not include, can play an important role in 
dissociation of halo nuclei owing to their small binding energies \cite{Ogata03,Ogata06}.

Although with the new generation computing facilities, 
the exact calculations for the breakup of the weakly-bound or halo nuclei are 
feasible, still some experimentalists insist on using the 
VPT \cite{Nakamura17,Rahman17,Cook20}. Also, in the breakup of these systems, 
the role of nuclear interactions, CNI effects, and multistep effects which VPT misses out 
is yet to be fully understood at intermediate and higher incident energies. 
In order to address the role of these factors at these energies, we aim to carry out 
the systematic investigation on the breakup of various two-body projectiles using the continuum-discretized coupled-channels method 
(CDCC) \cite{Yahiro12}. The choice of CDCC is based on the fact that it is a quantum mechanical reaction model, which
treats breakup processes of all-orders by Coulomb and nuclear components as well as their interference terms on equal footing. 
For efficiency, the eikonal version of CDCC (E-CDCC) \cite{Ogata03,Ogata06}, is employed in the numerical calculations.

In this work, we study the breakup of 
two different two-body weakly-bound projectiles, 
$^{11}$Be ($s$-wave) 
having a $\rm C$-$n$ structure and 
$^{17}$F ($d$-wave) having a $\rm C$-$p$ structure, on a heavy 
$^{208}$Pb target.
We choose three different incident energies $520$\,MeV/A (GSI energy), 
$250$\,MeV/A (RIKEN energy), and a lower energy $100$\,MeV/nucleon.

This paper is organized as follows. 
In Section~\ref{S2}, we briefly described the formulation of the E-CDCC. 
In Section~\ref{MS}, we tabulated various model parameters used in the present study. 
In Sections~\ref{Core-n} and \ref{core-p}, we discussed our main results for 
the breakup of ${\rm C}$-$n$ structure projectile and ${\rm C}$-$p$ structure projectile, respectively. 
Finally, we gave the conclusions in Section~\ref{Sum}.

\section{Model Formulation}
\label{S2}
\begin{figure}[htbp]
\begin{center}
\includegraphics[width=7cm]{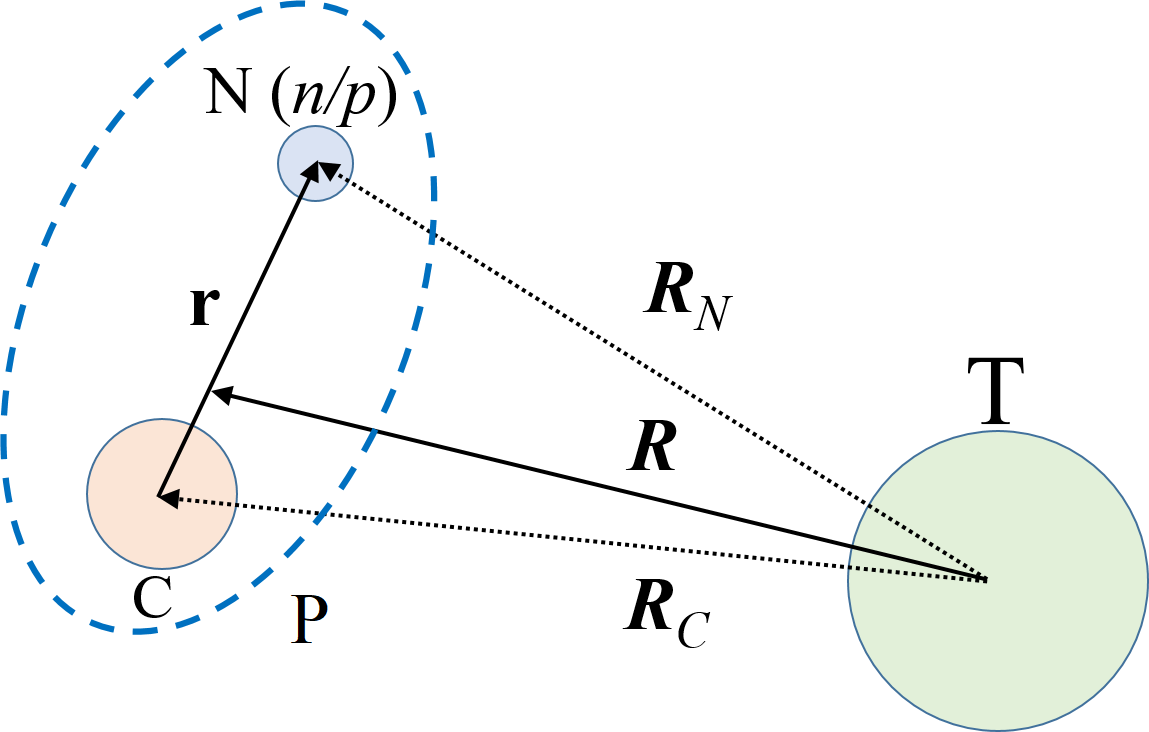}
\caption{Schematic illustration of the three-body (${\rm T}+{\rm C}+{\rm N}$) system.}
\label{Fig1}
\end{center}
\end{figure}

In CDCC, the projectile (${\rm P}$) is considered to be composed of two particles, 
the core (${\rm C}$) and a valence nucleon (${\rm N}=n$ or $p$). The scattering of ${\rm P}$ on target (${\rm T}$), 
is described by a three-body (${\rm T}+{\rm C}+{\rm N}$) model, schematically shown in Fig.~\ref{Fig1}. 
The coordinates of ${\rm P}$, ${\rm C}$, and ${\rm N}$ relative to ${\rm T}$ are represented by 
${\bm R}$, ${\bm R}_{\rm C}$, and ${\bm R}_N$, respectively, and ${\bm r}$ represents
the coordinate from ${\rm C}$ to $\rm N$. The spin of ${\rm C}$ and ${\rm T}$ are neglected in the present study. 
The three-body Schr\"odinger equation is given by
\begin{equation}
\left[  -\frac{\hbar^{2}}{2\mu}{\bm \nabla}_{{\bm R}}^{2}%
+U_{\mathrm{C}}(  R_{\mathrm{C}})+U_{N}(  R_{N})  
+\hat{h}-E\right]  \Psi(  {\bm r},{\bm R})  =0,
\label{sch}%
\end{equation}
where $\mu$ is the reduced mass of the ${\rm P}$-${\rm T}$ system, $\hat{h}$ is the internal Hamiltonian of ${\rm P}$, $E$ is the total energy of the system, 
and $U_k(k={\rm C}, {\rm N})$ is the interaction between  ${\rm T}$ and each constituent of ${\rm P}$ 
containing both nuclear and Coulomb components. 
Note that, for ${\rm N}=n$, ${\rm P}$-${\rm T}$ interaction contains only the nuclear part. 

In CDCC, the wave function of the reaction system is expanded in terms of the eigenstates, 
including bound and continuum states, of $\hat{h}$. 
To discretize the continuum, the average method in which the continuum states within each bin are averaged 
into a single state, is adopted and given by
\begin{equation}
 \phi_{i}(r)=\dfrac{1}{\sqrt{\Delta k}}\int_{k_{i}}^{k_{i+1}} \phi_{k}(r) dk, 
\end{equation}
where $k$ is the C-N relative wave number, and it is divided into a bin of size $\Delta k =k_{i+1}-k_i$. 

In E-CDCC \cite{Ogata03,Ogata06}, the total wave function $\Psi(  {\bm r},{\bm R})$
is described by
\begin{equation}
\Psi(  {\bm r},{\bm R})  =
\sum_{i}\frac{1}{\sqrt{\hbar v_{i}}}
e^{i(K_{i}z+\eta_i \ln (K_i R-K_i z))}
\psi_{i}(  b,z)
\phi_{i}({\bm r}),
\label{psi-ecdcc}
\end{equation}%
where $\phi_{i}({\bm r})$ is the wave function of ${\rm P}$ in the
ground state ($i=0$) and discretized continuum states 
($i\neq0$) satisfying
$\hat{h}\phi_{i}(  {\bm r})  =\varepsilon_{i}\phi_{i}({\bm r})$,
$K_{i}$ ($v_i$) is the relative wave number (velocity) between ${\rm P}$ and ${\rm T}$, 
$\eta_i$ is the Sommerfeld parameter, $b$ is the impact parameter, and $\phi_R$
is the azimuthal angle of ${\bm R}$. The $z$-axis is taken to be the
incident direction.
For simplicity, in Eq.~(\ref{psi-ecdcc}), the $\phi_R$
dependence of the wave function is dropped.
It should be noted that the monopole Coulomb interaction
between P and A is taken into account, by using the Coulomb incident wave
function in Eq.~(\ref{psi-ecdcc}).

The set of coupled-channel equations derived from Eqs.~~(\ref{sch}) and (\ref{psi-ecdcc}) 
are solved and the breakup cross sections are computed from the scattering matrix 
(S-matrix) obtained. 
\section{Results and Discussion}
\label{RAD}
\subsection{Model setting}
\label{MS}

Table~\ref{T1} shows the radius parameter $r_0$, the diffuseness parameter $a_0$, the relative angular momentum 
$\ell$ of the ${\rm C}$-$n$ or ${\rm C}$-$p$ pair in the ground state, $\ell_0$, together with one neutron or proton 
separation energy $S_{n/p}$ and the number of bound states ($n_b$).

\begin{table}[h]
\caption{Parameters adopted for core (${\rm C}$) and a valence nucleon (${\rm N}=n$ or $p$) pair.}
\centering
\begin{tabular}{ccccccc}
\hline\hline\\[-1.5ex]
System & $r_0$\,(fm)     & $a_0$\,(fm) & $\ell_0$     & $S_{n/p}$\,(MeV)& $n_b$&Ref. \\ [1ex]
\hline\hline\\[-1.5ex]
$^{11}$Be & 1.20 & 0.60         & 0         & 0.503 &2& \cite{Capel04}\\
$^{17}$F  & 1.20 & 0.64         & 2         & 0.600 &2& \cite{Spa00}\\[1ex]
\hline\hline
\end{tabular}
\label{T1}
\end{table}
The depths of the central and spin-orbit (LS) potential for $^{11}$Be are adopted 
from Table-I of Ref.~\cite{Capel04}, to reproduce the major-low lying states (see Table-II of Ref.~\cite{Capel04}) 
of $^{11}$Be. However, the bound excited state of $^{17}$F is the $s$-wave, due to which the effect of the spin of the external 
proton and the LS potential are found to be negligible on breakup and inelastic cross sections. 
Thus, in the present study, the spin of the external proton is neglected for $^{17}$F. The depths of the central 
potential for the $d$- and $s$-waves are determined to reproduce the energies of the ground state and the bound 
excited state at $0.105$\,MeV, respectively. The depth parameters of $52.10$\,MeV ($48.00$\,MeV) was adopted 
to obtain the non-resonant continuum for the $p$-wave ($f$-wave). 
The maximum value $\ell_{\rm max}$ of $\ell$ is set to $3$ for all cases, and 
the maximum value of $r$ is chosen to be $200$\,fm.
The ${\rm C}$-$n/p$ relative wave number $k$ is discretized for each $\ell$, by the 
momentum-bin method with an equal increment $\Delta k$. 
Table~\ref{T2} shows the converged values of $k$ and $\Delta k$ for each case.

\begin{table}[h]
\caption{Converged values of model parameters, $k$ and $\Delta k$ for each case.}
\centering
\begin{tabular}{cccc}
\hline\hline\\[-1.5ex]
System & Energy&$k$     & $\Delta k$ \\ 
&(MeV/A)&(fm$^{-1}$)     & (fm$^{-1}$) \\ [1ex]
\hline\hline\\[-1.5ex]
             &100 & 0.78 & 0.040    \\
$^{10}$Be$+n$&250 & 0.77 & 0.044     \\
             &520 & 0.77 & 0.044     \\ [1ex]
\hline\\[-1.5ex]
             &100 & 1.26 & 0.072    \\
$^{16}$O$+p$ &250 & 1.26 & 0.072      \\
             &520 & 1.26 & 0.072     \\ [1ex]
\hline\hline
\end{tabular}
\label{T2}
\end{table}

\begin{figure}[H]
\begin{center}
\includegraphics[width=7cm]{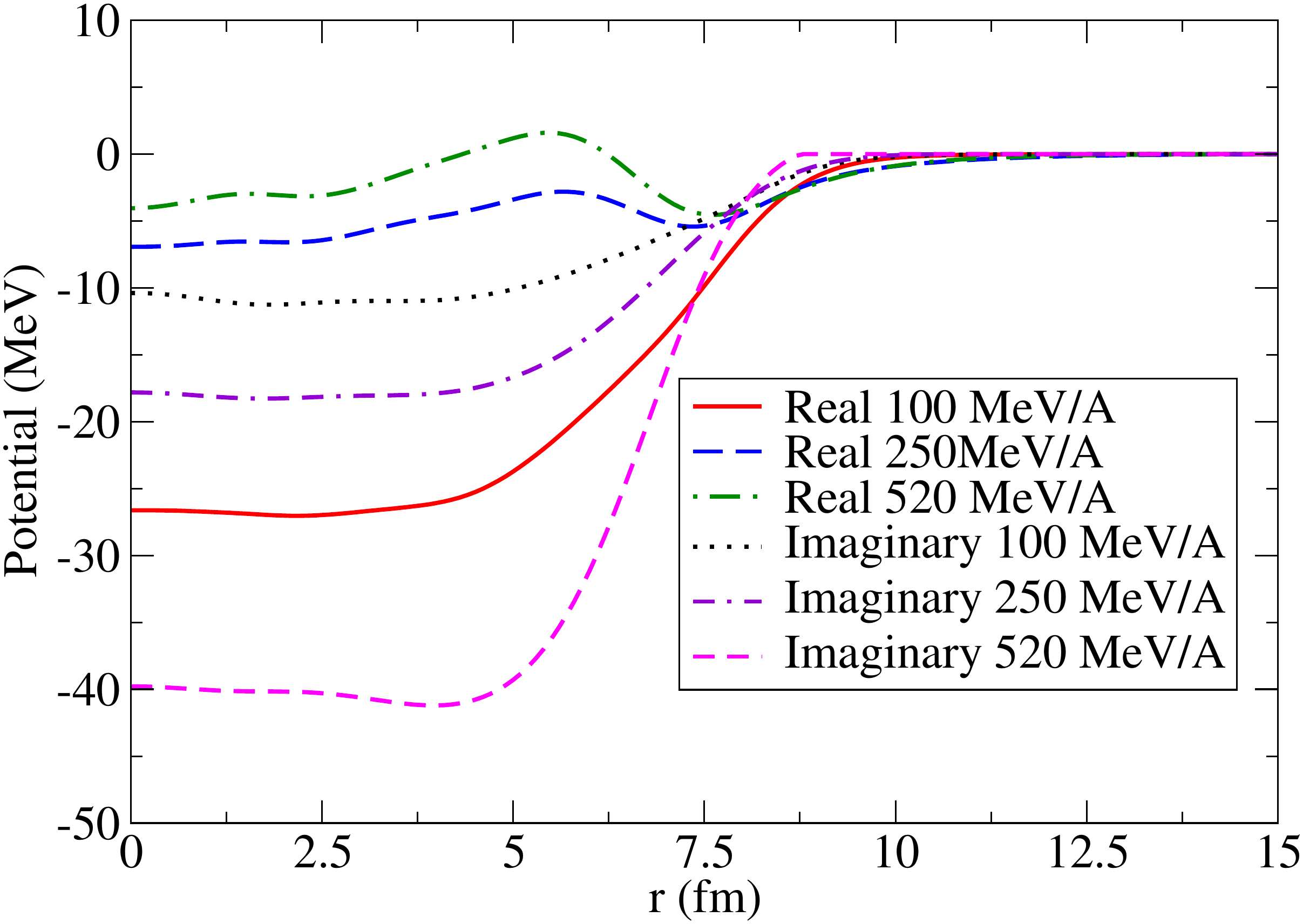}
\caption{Real and imaginary components of the neutron-Target ($n$-T) potential at different energies.}
\label{Fig2}
\end{center}
\end{figure}

The distorting nucleon-nucleus (${\rm N}$-${\rm T}$) and nucleus-nucleus (${\rm C}$-${\rm T}$) potentials 
are evaluated by a microscopic folding model.
The Melbourne nucleon-nucleon $g$ matrix~\cite{Amo00} and
the Hartree-Fock wave functions of ${\rm C}$ and ${\rm T}$ based on the Gogny D1S force~\cite{DG80,Ber91}
are adopted. This microscopic approach has successfully been applied to
several reaction systems~\cite{Min12,Yahiro12,Sum12}.
The maximum impact parameter $b_{\rm max}$ is set to be
$50$~fm for nuclear breakup processes, whereas we put
$b_{\rm max}=700$~fm when Coulomb breakup is included. 
The real and imaginary potentials of $n$-${\rm T}$ at three energies is shown in Fig.~\ref{Fig2}. 
We observe that the real part of $n$-${\rm T}$ interaction is strongly attractive at $100$\,MeV/A, 
attractive but shallow at $250$\,MeV/A, and slightly repulsive at $520$\,MeV/A.
On the other hand, the imaginary part is deepest at $520$\,MeV/A and shallow at $100$ and $250$\,MeV/A. 
The same trend is found for the ${\rm C}$-${\rm T}$ potentials. 
The numerical data of the optical potentials used in this study can be provided on request via e-mail.

\subsection{Breakup of core-neutron projectile ($^{11}$Be)}
\label{Core-n}
In this section, we present our results for 
C-$n$ projectile $^{11}$Be ($s$-wave, light-mass system) 
breakup on $^{208}$Pb target 
at three different choices of the beam energy.

\begin{figure}[ht]
\begin{center}
\includegraphics[width=7.1cm]{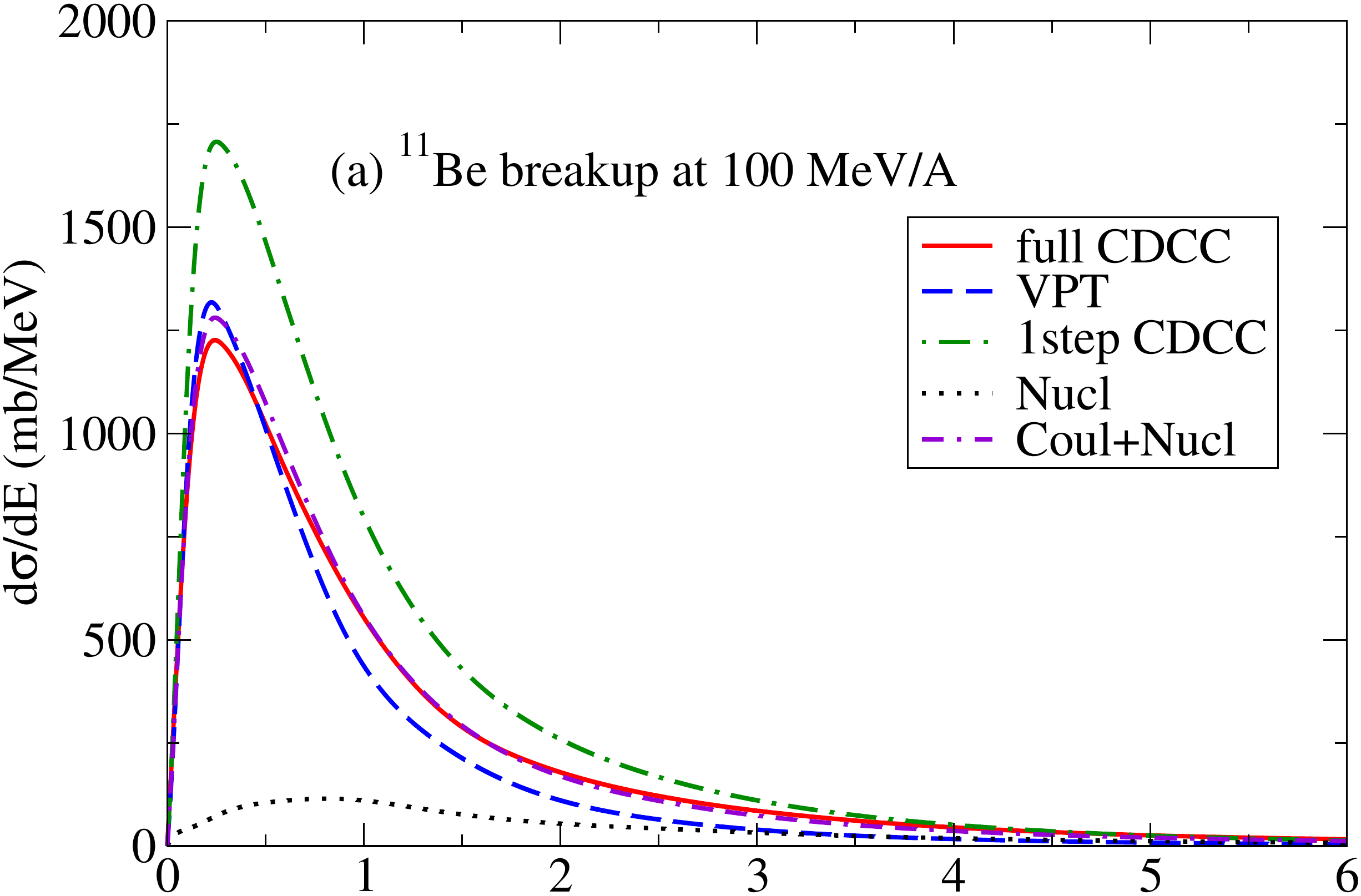}
\includegraphics[width=7.1cm]{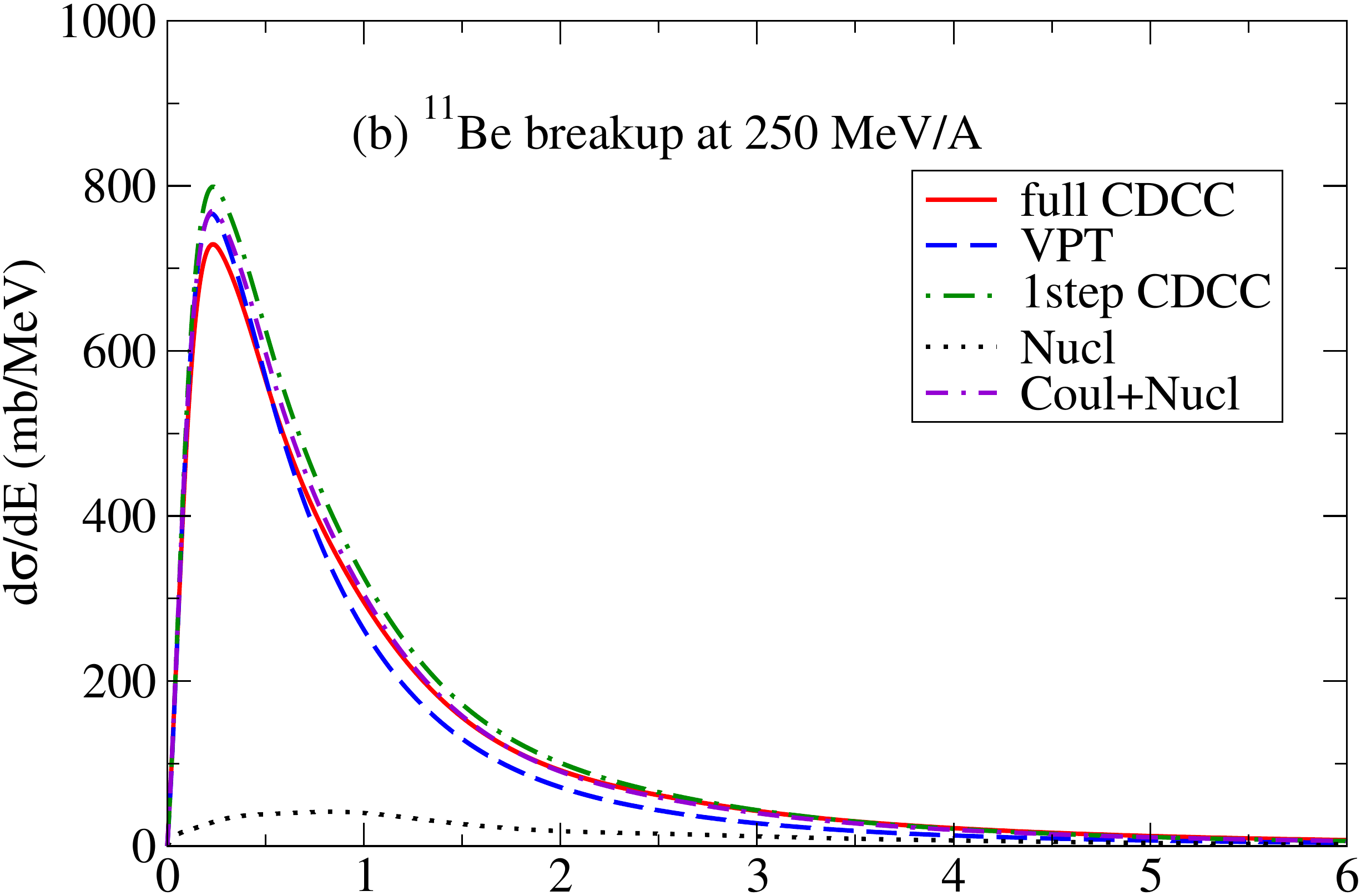}
\includegraphics[width=7.1cm]{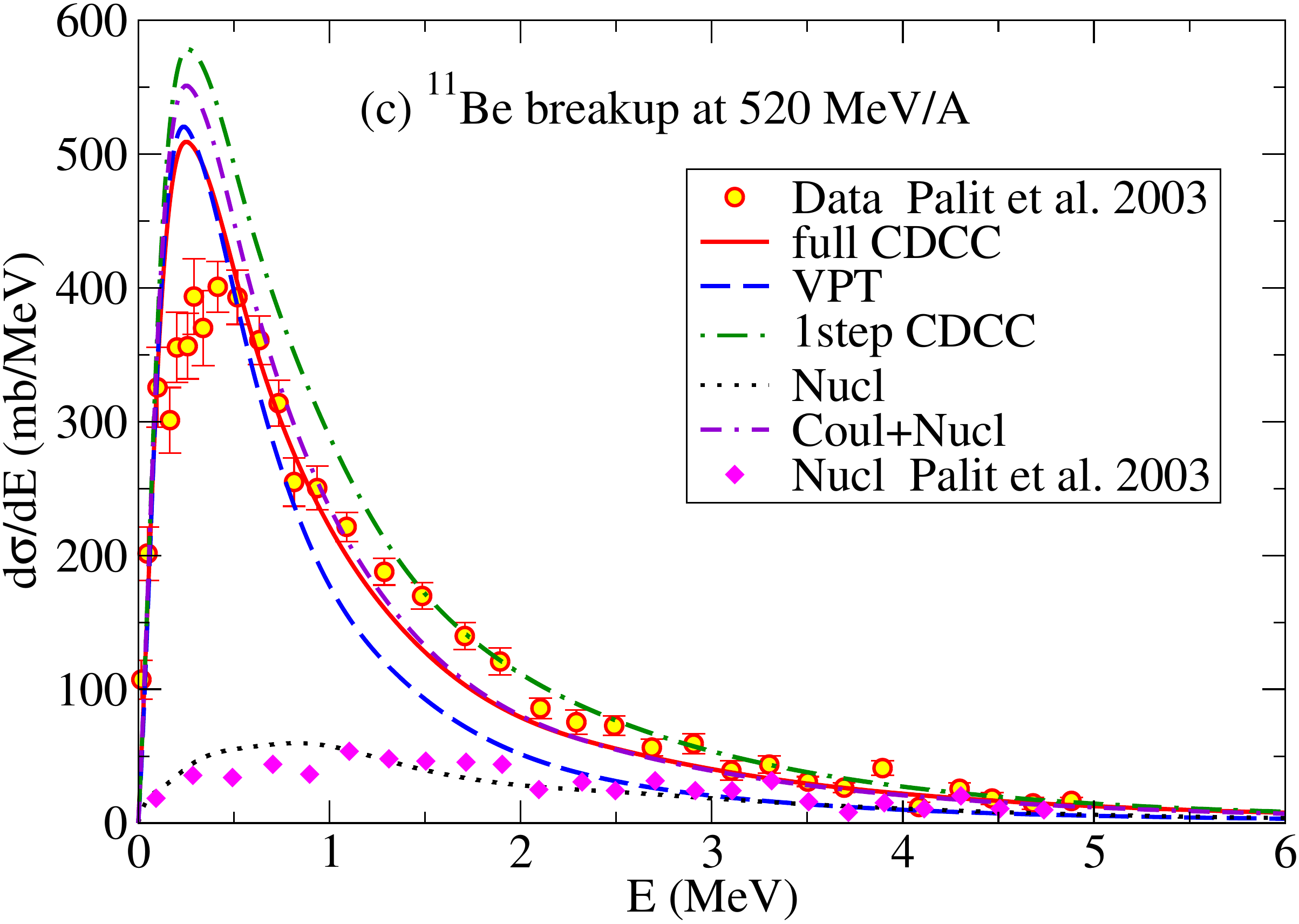}
\caption{Breakup cross section of $^{11}$Be on $^{208}$Pb target at 
(a) $100$\,MeV/A, (b) $250$\,MeV/A, and (c) $520$\,MeV/A,  
plotted as a function of relative energy $E$ between the $^{10}$Be core and the valence neutron after dissociation. The solid, 
dashed, dot-dashed, dotted, and dash-dotted lines correspond to full CDCC, VPT, 
1step CDCC, nuclear breakup, and incoherent sum of the Coulomb and nuclear breakup calculations, respectively. 
For details refer to text. Experimental data are adopted from the Ref.~\cite{Palit03}.}
\label{fig3}
\end{center}
\end{figure}

The $^{11}$Be breakup cross sections as a function of the $^{10}$Be-$n$ relative energy $E$ 
are presented in Fig.~\ref{fig3};~(a),~(b),~and~(c) correspond to $100$\,MeV/A, $250$\,MeV/A, and $520$\,MeV/A, respectively. 
The various results shown in each panel of Fig.~\ref{fig3} consist of the following settings:
(i) The full CDCC (solid line) calculation, which refers to the results with both the Coulomb and 
nuclear breakup including the CNI, and the contribution from multistep processes. 
(ii) A calculation mimicking that of VPT (dashed line), 
which correspond to the first-order Coulomb breakup restricted to the $E1$ contribution and with no nuclear breakup contribution. 
(iii) One-step CDCC (dot-dashed line) meaning the first-order perturbative calculations, which ignores the contribution 
from multistep processes (without continuum-continuum and back couplings). 
(iv) Nuclear breakup (dotted line) calculation, which refers to the contribution obtained by switching off the 
Coulomb breakup amplitudes.  It is worth mentioning that, the definition of nuclear and Coulomb 
components depends upon the various conditions and is an interesting theoretical subject as discussed 
in Ref.~\cite{Desco17}; however, it is beyond the scope of the present study. 
(v) The incoherent sum of the full Coulomb breakup (Coul), in which the multistep Coulomb breakup with all multipoles is included without the nuclear breakup, 
and full nuclear breakup (Nucl) (Coul$+$Nucl, dash-dotted), is plotted to study the CNI effects.
 
Now, let us turn our focus to $^{11}$Be breakup on $^{208}$Pb at $520$\,MeV/A in Fig.~\ref{fig3}~(c).
It can be clearly seen that the VPT agrees with the full CDCC calculation around the peak, however, a
difference originating from the missing nuclear breakup contribution and multistep processes in VPT appears 
in the tail region of the cross section.

The good agreement in the peak region seems to confirm the success of the VPT calculation. However, 
1step CDCC gives a significantly larger cross section than full CDCC in the same region as shown in 
Fig.~\ref{fig3}~(c), which indicates the failure of the VPT. 

\begin{figure}[ht]
\begin{center}
\includegraphics[width=7.1cm,clip=]{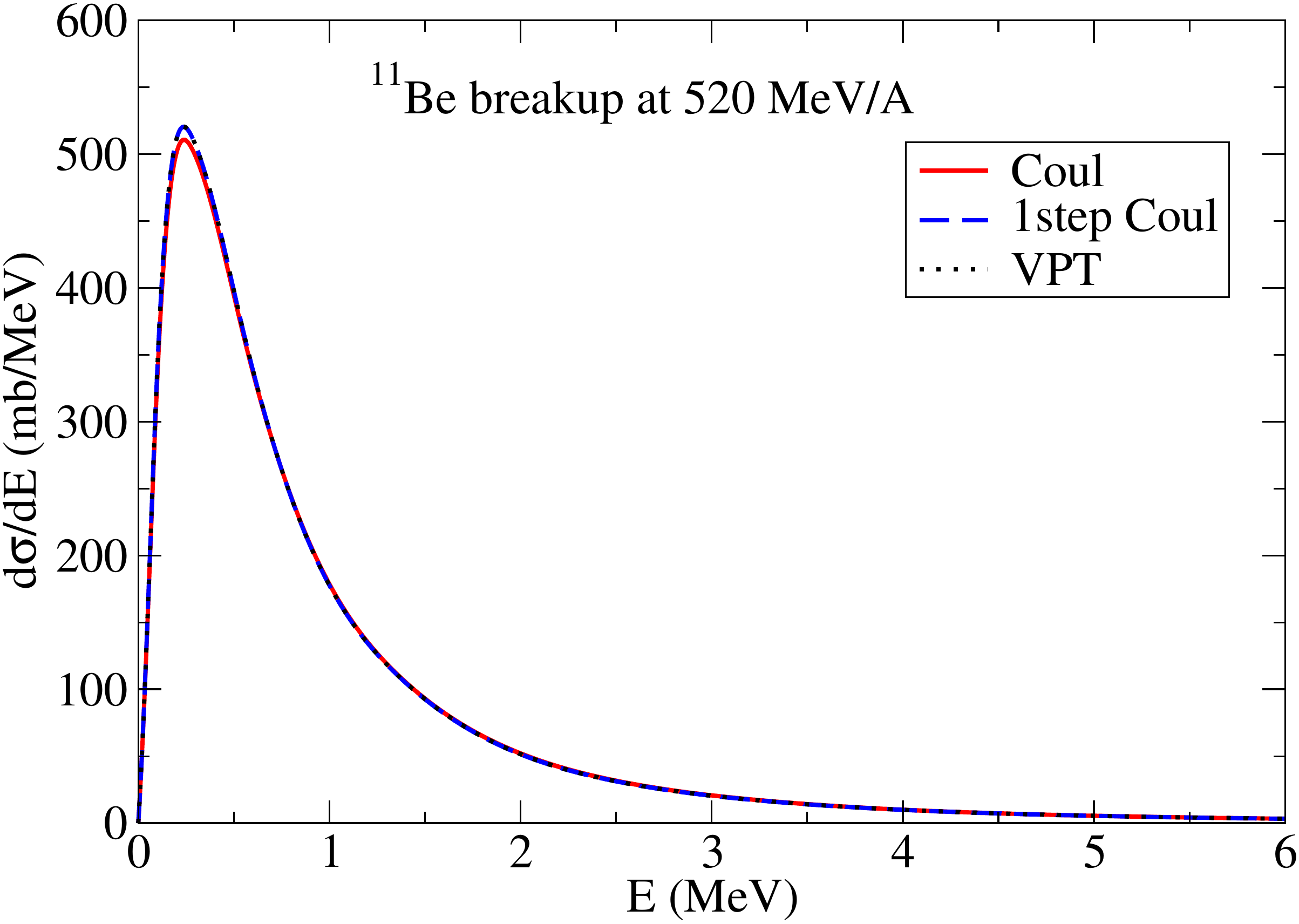}
\caption{The Coul, 1step Coul and VPT breakup cross section of $^{11}$Be on $^{208}$Pb target at $520$\,MeV/A, plotted as a function of relative energy $E$ between 
the $^{10}$Be core and the valence neutron after dissociation. 1step Coul is the same as Coul, except that multistep breakup is disregarded.}
\label{fig3a}
\end{center}
\end{figure}

\begin{figure}[ht]
\begin{center}
\includegraphics[width=7.1cm,clip=]{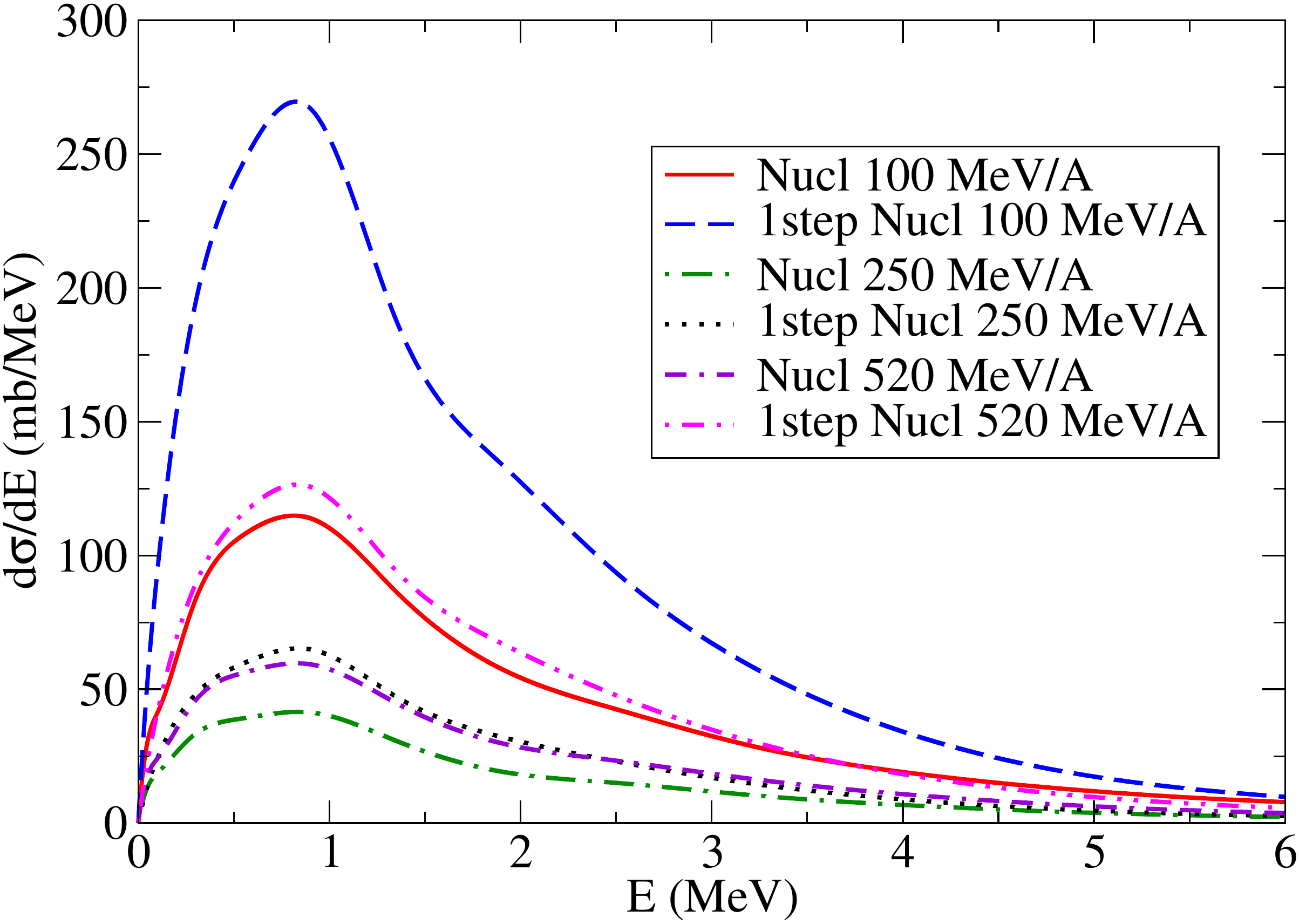}
\caption{The Nucl and 1step Nucl breakup cross section of $^{11}$Be on $^{208}$Pb target at $100$\,MeV/A, $250$\,MeV/A, 
and $520$\,MeV/A, plotted as a function of relative energy $E$ between 
the $^{10}$Be core and the valence neutron after dissociation. 
1step Nucl refers to the Nucl calculation without the multistep breakup.}
\label{fig3b}
\end{center}
\end{figure}

To understand the situation more clearly, we first confirmed that the multistep effect on Coulomb breakup is negligible, 
whereas that on nuclear breakup is very significant (shown in Fig.~\ref{fig3a} and \ref{fig3b}). 
This indicates that the large difference between 1step CDCC 
and full CDCC is as a result of the multistep nuclear breakup processes. 

Physically, the Coulomb and nuclear breakup contributions should sum up coherently 
and depending upon the situation it can lead to constructive or destructive CNI effect  
(refer to the discussion of Table~\ref{T3}). It should be noted that 
the constructive or destructive CNI effect is based on the total integrated breakup 
cross section values.  For reader's illustration, the unphysical 
incoherent sum of the full Coulomb and Nuclear breakup contributions in the breakup distributions 
are also shown in Fig.~\ref{fig3}.
For the CNI effect, we observe that the difference between the solid and dash-dotted lines is 
small but not negligible around the peak. It will be interesting if the CNI effect accidentally makes
the VPT result agrees with that of full CDCC.

Thus, the nuclear breakup component, the CNI effect, and the multistep breakup processes for the nuclear part are all found to
be non-negligible. We then conclude that the assumptions adopted in the VPT are not valid, even though the cross section 
obtained with the VPT agrees with that of full CDCC in the peak region. 
Although, our main purpose is not to reproduce the experimental data, Fig.~\ref{fig3}~(c) 
shows the good agreement of our calculations with the experimental data for the total and nuclear 
breakup contributions \cite{Palit03}.  

The same features can be seen at lower incident energies in Fig.~\ref{fig3}~(a) and (b). 
At $100$\,MeV/A, the role of higher multipoles is not negligible, which makes Coul$+$Nucl close to CDCC result. 
It is found that the bound excited state in $^{11}$Be reduces the $p$-wave breakup cross section because the inelastic cross section to 
the state is quite large. However, the aforementioned features of the results are not affected by the existence of the bound excited state.

The significant multistep effect on nuclear breakup is shown in Fig.~\ref{fig3b}.
One can see that the difference due to multistep processes depends quantitatively on the incident energies; 
it is about $39\%$, $9\%$, and $24\%$ at $100$, $250$, and $520$\,MeV/A, respectively. This is due to the 
energy dependence of the C-T and $n$-T nuclear potentials. At $100$\,MeV/A, the real (imaginary) potential, 
$V$ ($W$), is very deep (shallow) as shown in Fig.~\ref{Fig2}, which gives a large contribution to the breakup cross section. 
At $250$ MeV/A, $W$ becomes slightly deeper, whereas $V$ becomes very shallow. 
This shrinks the size of the nuclear breakup cross section. 
On the other hand, at $520$ MeV/A, $W$ becomes significantly deep and $V$ remains small. 
It is known that the role of $W$ is not only for absorption, but it is also a source of breakup. 
In the transition potential, $V$ and $W$ play the same role. Thus, the deep $W$ gives a large amount of the breakup cross section, even though it gives 
stronger absorption. The energy dependence of the multistep effect on the nuclear breakup cross section 
can be understood in this way. 

\begin{figure}[ht]
\begin{center}
\includegraphics[width=7.1cm]{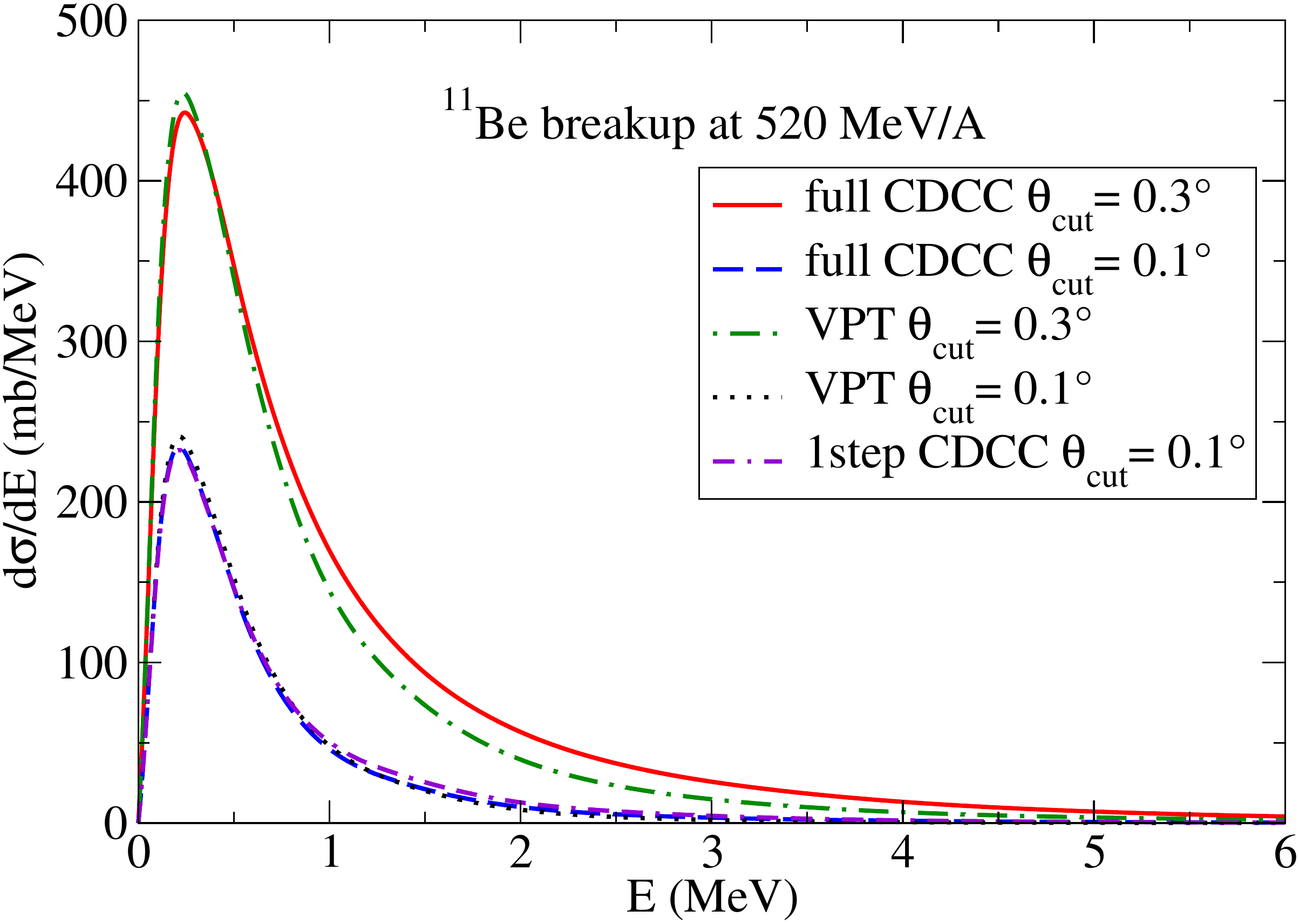}
\caption{Breakup cross section of $^{11}$Be on $^{208}$Pb target at $520$\,MeV/A. The full CDCC (solid and dashed lines), 
VPT (dot-dashed and dotted lines), and 1step CDCC (dash-dotted line) results for different choices of the cutoff of the scattering angle, $\theta_{cut}$, are plotted.}
\label{fig4}
\end{center}
\end{figure}

Now, we turn our attention to the dependence of the breakup cross section on the scattering angle $\theta$ 
of the core-nucleon center of mass system. 
Usually, experimentalists choose some cut on the scattering angle $\theta$, assuming that 
below that cutoff angle $\theta_{\rm cut}$, the nuclear breakup contribution can be ignored. 
In Fig.~\ref{fig4}, we present the comparison of full CDCC, VPT and 1step CDCC results for different choices of $\theta_{\rm cut}$ at $520$\,MeV/A. 
It can be clearly seen that with $\theta_{\rm cut}=0.3\degree$, which corresponds to a rather strong restriction of 
$\theta$, the results of full CDCC and VPT are still differ in the tail region. 
At $\theta_{\rm cut}=0.1^\circ$, the difference becomes negligible and the result of 1step CDCC also agree well with the two.
Our findings show that the choice of $\theta_{\rm cut}$ should be investigated carefully, depending on the system and beam energy.

\begin{table}[htb]
\caption{The integrated breakup cross sections in millibarn (mb) for $^{11}$Be 
breakup on $^{208}$Pb target at $100$, $250$ and $520$\,MeV/A. 
The numerical integration is performed up to $6$ MeV.}
\centering
\begin{tabular}{cccccc}
\hline\hline\\[-1.5ex]
Setting        &total          &$s$     &$p$       &$d$      &$f$ \\[1ex]
\hline\hline\\[-2.0ex]
&\multicolumn{5}{c}{\underline{$^{11}$Be $+$ $^{208}$Pb at 100 MeV/A}}\\[1ex]
full CDCC&1442.23&93.86&1101.68&184.48&62.22\\
1step CDCC&2011.09 ($+$39\%)&77.75&1625.08&198.20&110.06\\
VPT&1209.58 ($-$16\%)&0.003&1209.58&0.00&0.00\\
Coul&1179.42 ($-$18\%)&31.92&1099.30&46.06&2.14\\
1step Coul&1207.26 ($-$16\%)&0.003&1206.23&1.02&0.01\\
Nucl&261.92&40.04&73.32&101.45&47.11\\
1step Nucl&568.355&77.68&169.12&212.23&109.33\\
Coul+Nucl&1441.34 ($-$1\%)&71.96&1172.62&147.51&49.25\\
\hline\hline\\[-2.0ex]
\\ [-2.0ex]
&\multicolumn{5}{c}{\underline{$^{11}$Be $+$ $^{208}$Pb at 250 MeV/A}}\\
full CDCC&783.97&26.82&683.07&57.39&16.70\\
1step CDCC&851.87 ($+$9\%)&19.36&755.74&51.57&25.19\\
VPT&716.43 ($-$9\%)&0.003&716.43&0.00&0.00\\
Coul&710.25 ($-$9\%)&8.29&690.21&11.45&0.30\\
1step Coul&717.32 ($-$9\%)&0.003&716.81&0.51&0.003\\
Nucl&93.84&14.75&26.12&38.57&14.39\\
1step Nucl&140.64&19.37&41.71&54.39&25.16\\
Coul$+$Nucl&804.09 ($+$3\%)&23.04&716.33&50.02&14.69\\
\hline\hline\\[-2.0ex]
\\ [-2.0ex]
&\multicolumn{5}{c}{\underline{$^{11}$Be $+$ $^{208}$Pb at 520 MeV/A}}\\
full CDCC&604.69&26.41&493.05&61.67&23.44\\
1step CDCC&752.32 ($+$24\%)&34.61&565.87&96.07&55.76\\
VPT&498.39 ($-$18\%)&0.002&498.39&0.00&0.00\\
Coul&495.93 ($-$18\%)&3.38&487.90&4.57&0.09\\
1step Coul&498.20 ($-$18\%)&0.002& 497.88&0.32&0.002\\
Nucl&139.60&20.52&45.00&52.10&22.10\\
1step Nucl&280.28&34.63&92.71&97.10&55.84\\
Coul+Nucl&635.53 ($+$5\%)&23.90&532.90&56.67&22.19\\
\hline\hline
\end{tabular}
\label{T3}
\end{table}

In Table~\ref{T3}, we list the total and each partial-wave ($s$, $p$, $d$, and $f$) integrated breakup cross sections 
calculated with different settings. 
The positive/negative percentage deviation is shown for various settings with respect to the full CDCC calculation. 
The positive/negative deviation of Coul$+$Nucl is due to the constructive/destructive CNI effect.
For some settings, the deviations are same, but the corresponding distributions show small differences. 
It is noted that, the difference around peak in the distributions result in small deviation, whereas, the 
difference in tail result in large deviation.

It can be seen in Table~\ref{T3},  that CNI is destructive in the $p$-wave and weakly constructive
in all other partial-waves for $^{11}$Be at each incident energy. 
This feature can be understood as the transitions contributing to $E1$ breakup shows significant destructive CNI effect.
For the total Coul$+$Nucl, CNI is found to be negligibly destructive, weakly constructive, and constructive for $^{11}$Be, 
at $100$\,MeV/A,  $250$\,MeV/A, and $520$\,MeV/A, respectively. 
Thus, the CNI effect appearing from incoherent sum of Coul and Nucl is found to be weak but 
not negligible for $^{11}$Be.

We have also performed the same set of calculations for an another C-$n$ case, $^{31}$Ne, with different ground state 
configuration \textit{i.e.}, $p$-wave. Qualitatively, for the role of nuclear breakup contribution and multistep processes, 
we have observed the trends similar to $^{11}$Be. In summary, for C-$n$ projectiles, despite the 
different ground state configurations, all the higher-order processes which are missing in VPT are significant.

\subsection{Breakup of core-proton projectile ($^{17}$F)}
\label{core-p}
In this section, we present our results for the breakup of 
a C-$p$ projectile $^{17}$F ($d$-wave, medium-mass system) in the same manner as in the previous section.
\begin{figure*}[htbp]
\begin{center}
\includegraphics[width=7.1cm]{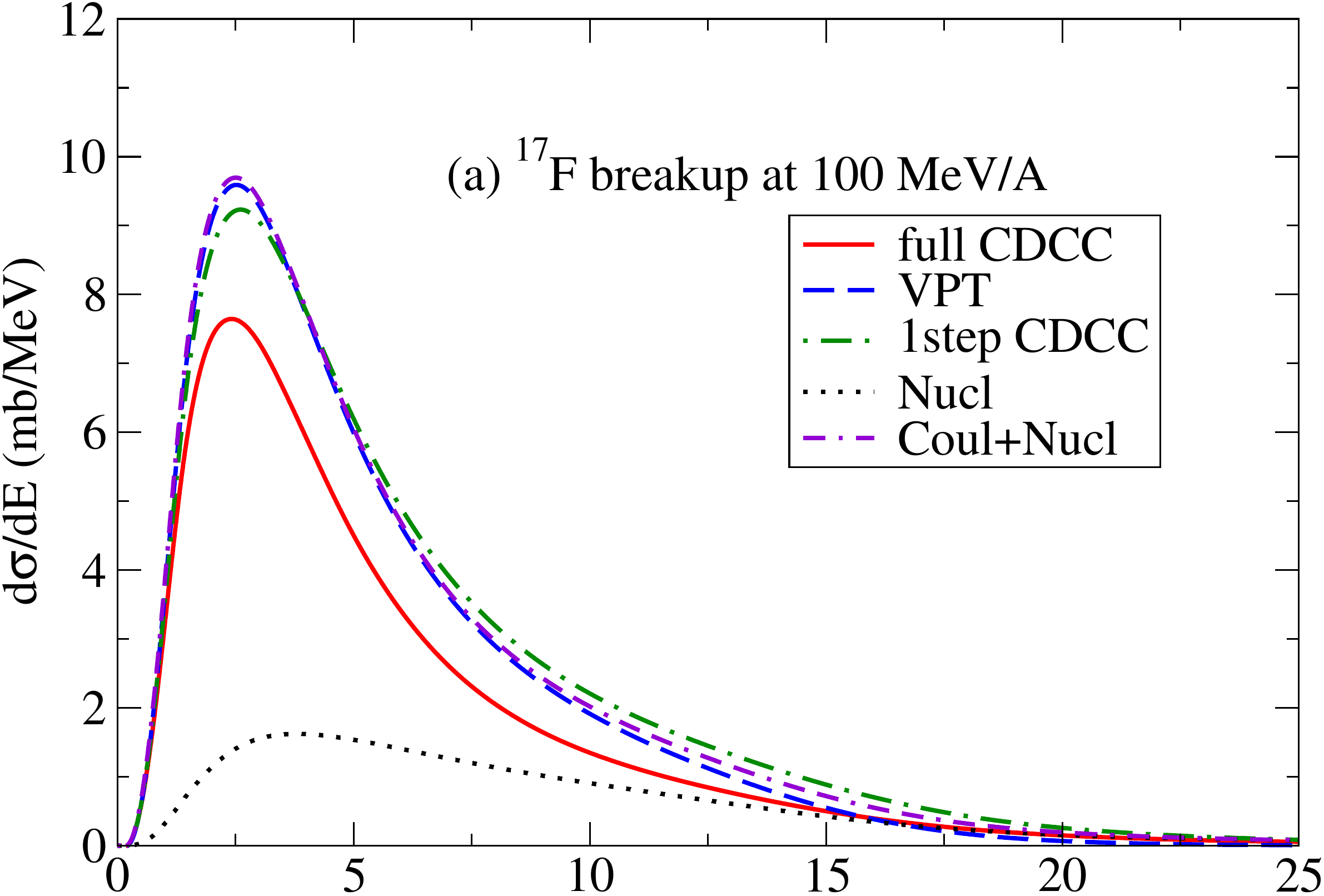}\hspace{20pt}
\includegraphics[width=7.1cm]{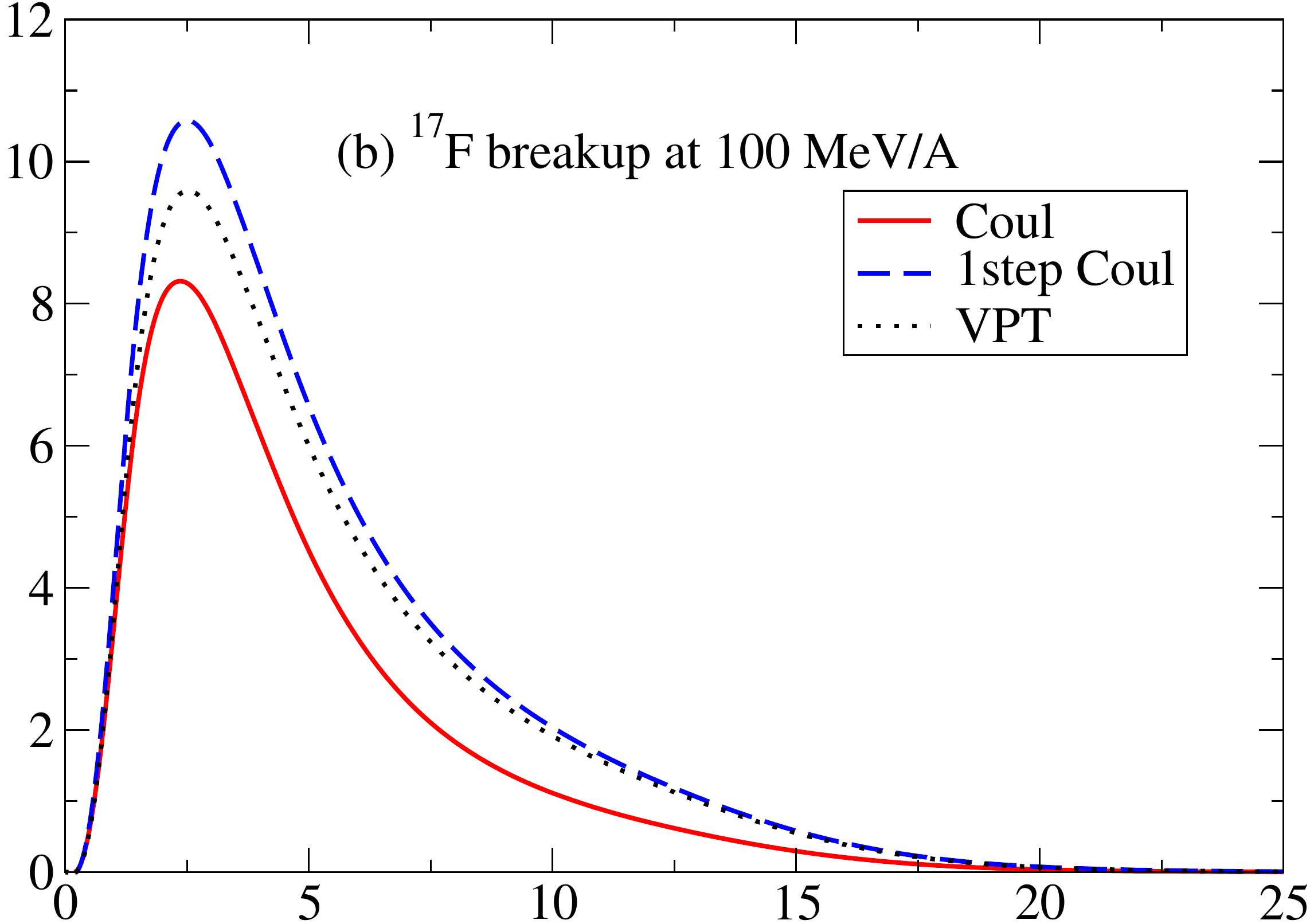}
\includegraphics[width=7.1cm]{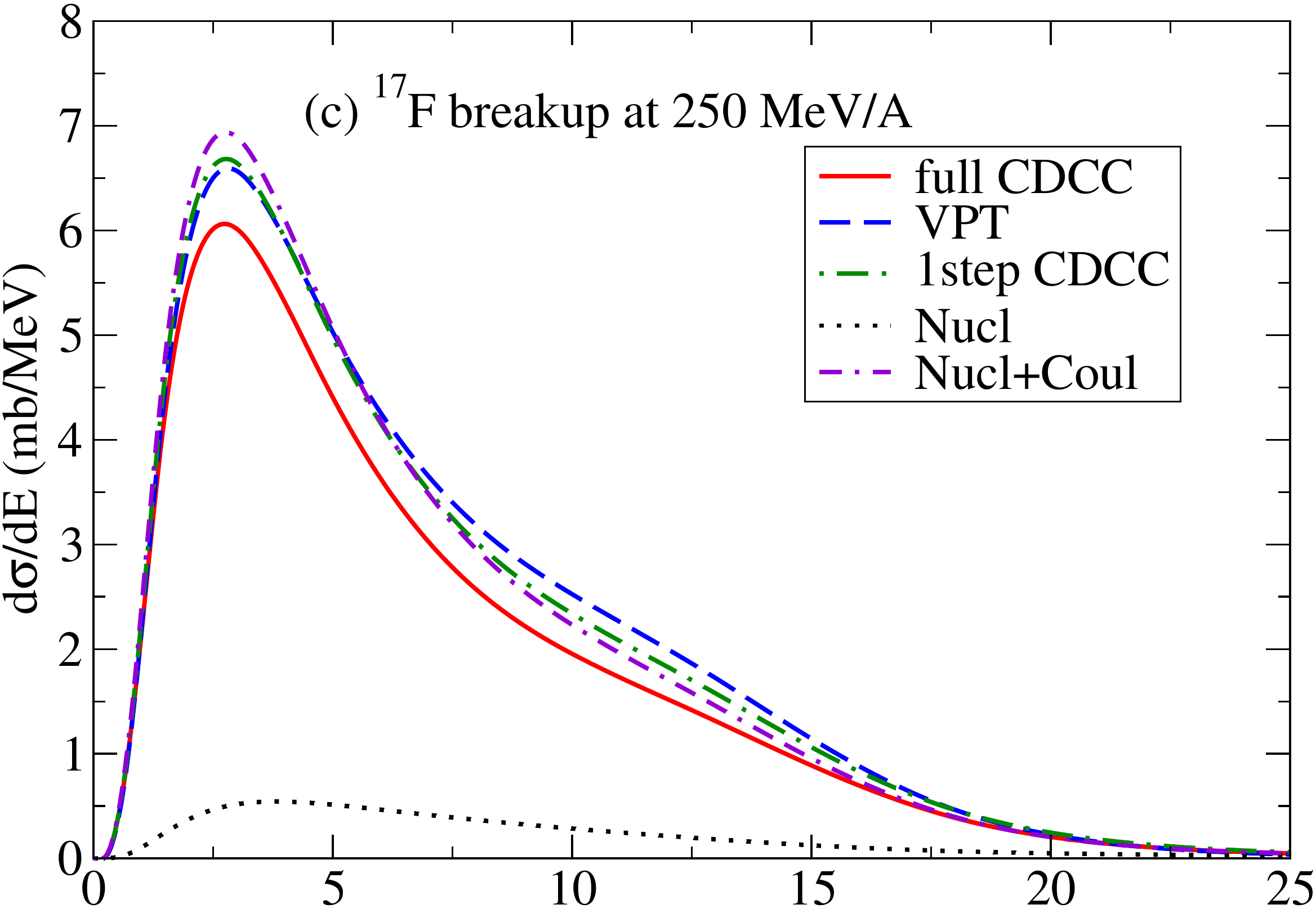}\hspace{20pt}
\includegraphics[width=7.1cm]{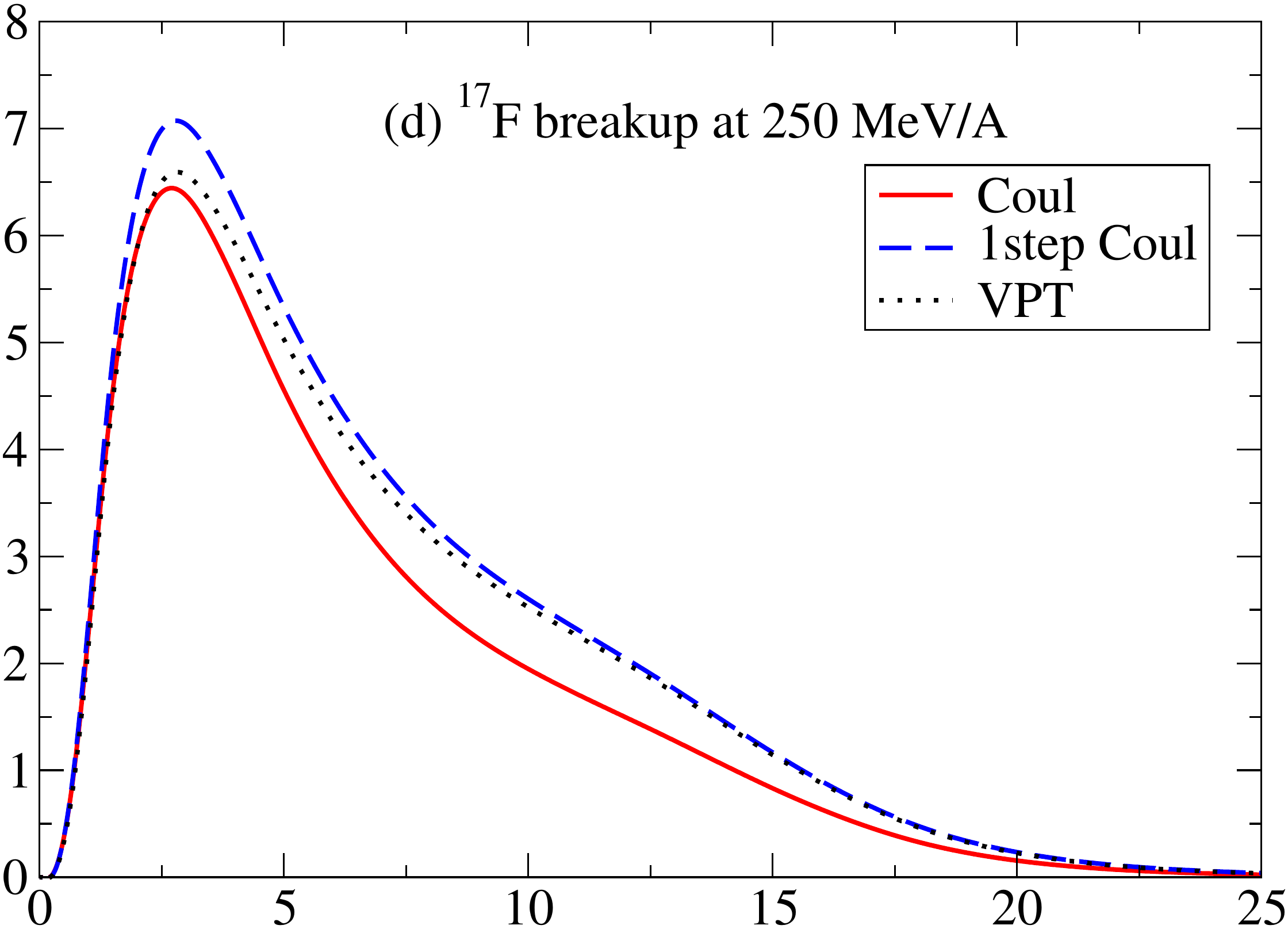}
\includegraphics[width=7.1cm]{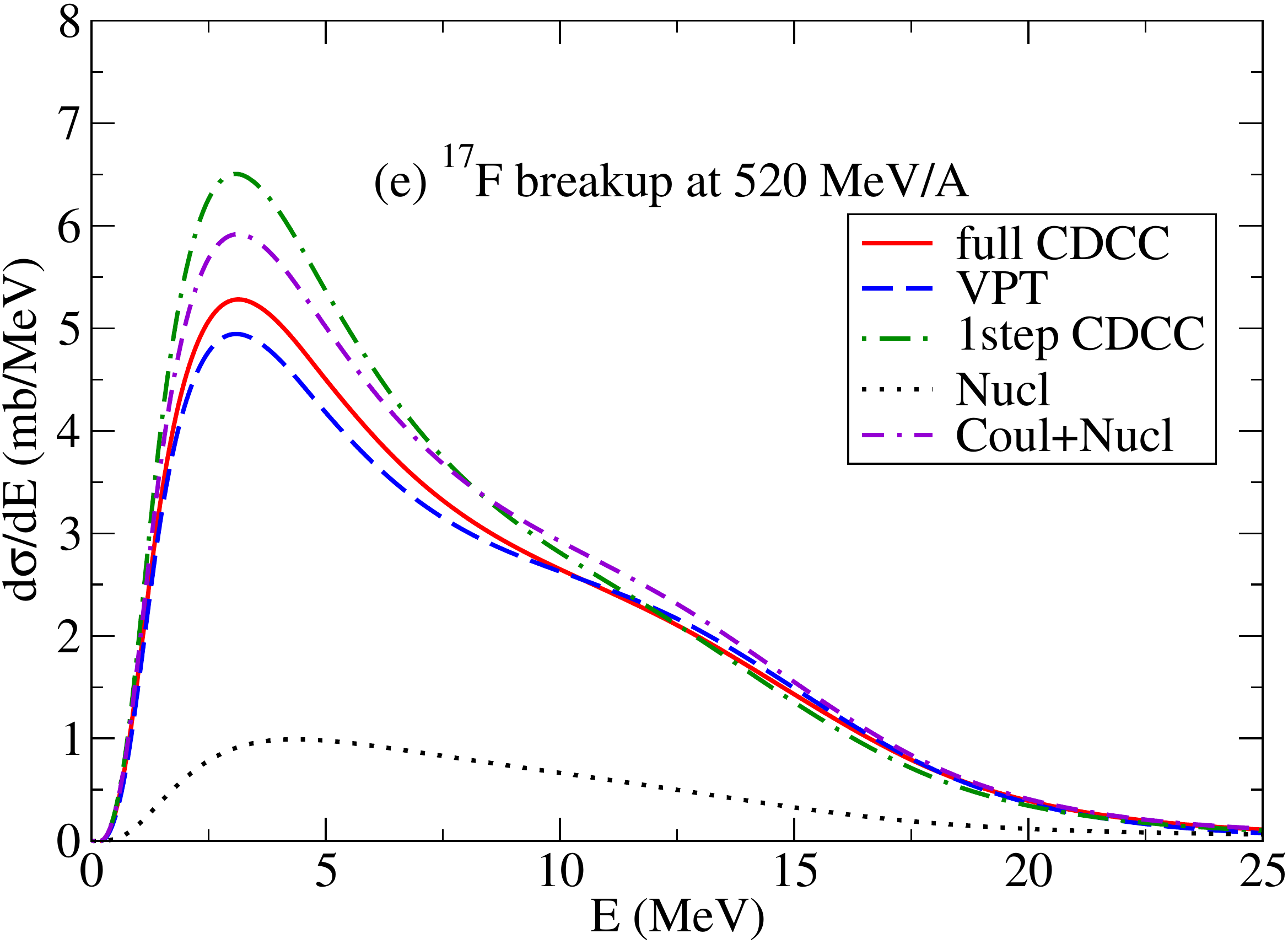}\hspace{20pt}
\includegraphics[width=7.1cm]{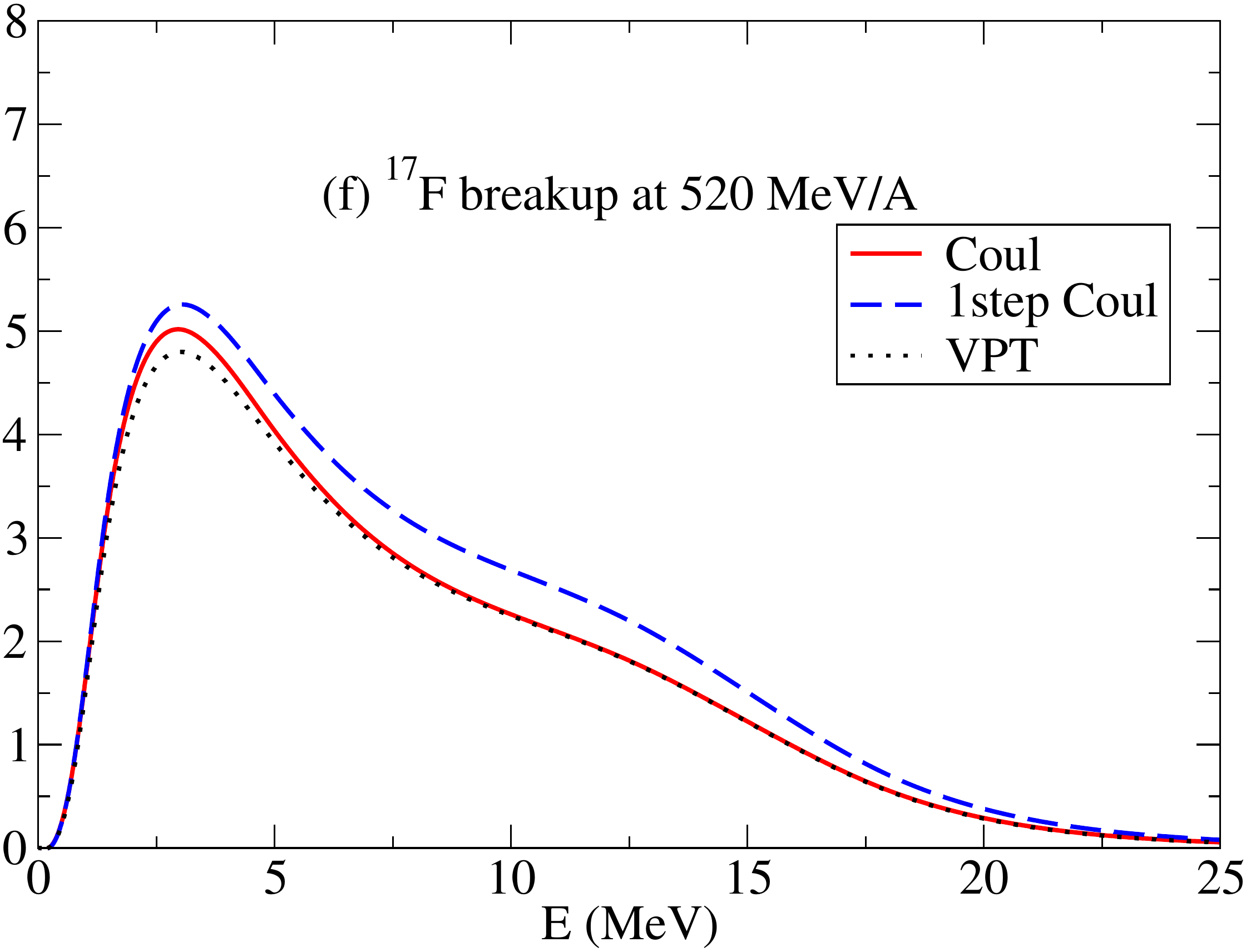}
\caption{Breakup cross section of $^{17}$F on $^{208}$Pb target, (a) and (b) at $100$\,MeV/A, 
(c) and (d) at $250$\,MeV/A, (e) and (f) at $520$\,MeV/A, plotted as a function of relative energy $E$ between 
the $^{16}$O core and the valence proton after dissociation. 
In panel (a), (c), and (e), the solid, dashed, dot-dashed, dotted, and dash-dotted lines correspond to full CDCC, VPT, 
1step CDCC, nuclear breakup, and incoherent sum of Coulomb and nuclear breakup calculations, respectively. In panel (b), (d), 
and (f), the solid, dashed, and dotted lines correspond to result of full Coulomb, 1step Coulomb, and VPT calculations, respectively. For details refer to text.}
\label{fig8}
\end{center}
\end{figure*}

\begin{figure}[htbp]
\begin{center}
\includegraphics[width=7.1cm,clip=]{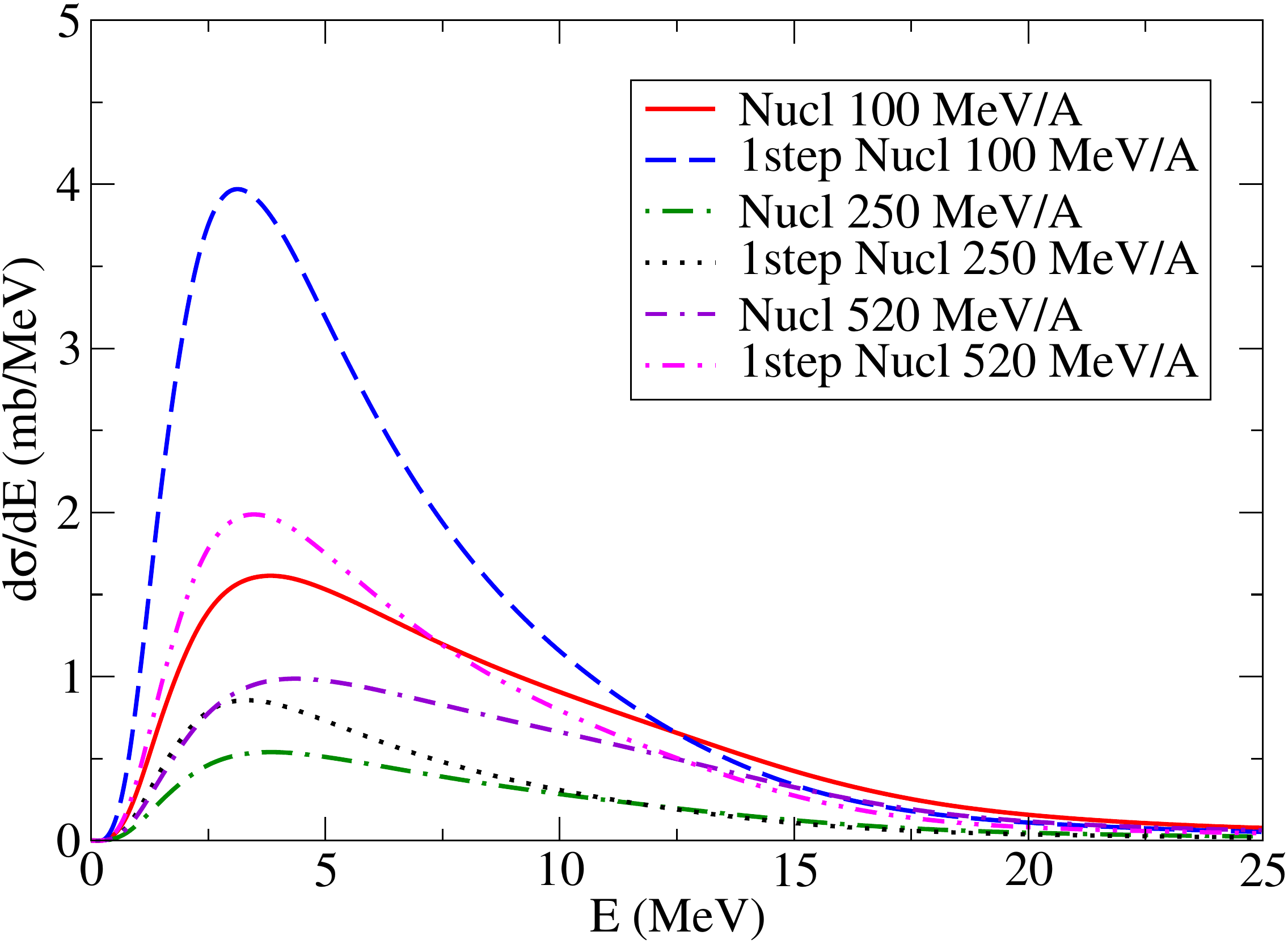}
\caption{The Nucl and 1step Nucl breakup cross section of $^{17}$F on $^{208}$Pb target at $100$\,MeV/A, $250$\,MeV/A, 
and $520$\,MeV/A, plotted as a function of relative energy $E$ between 
the $^{16}$O core and the valence proton after dissociation.}
\label{fig8N}
\end{center}
\end{figure} 
\begin{figure}[htbp]
\begin{center}
\includegraphics[width=7.1cm]{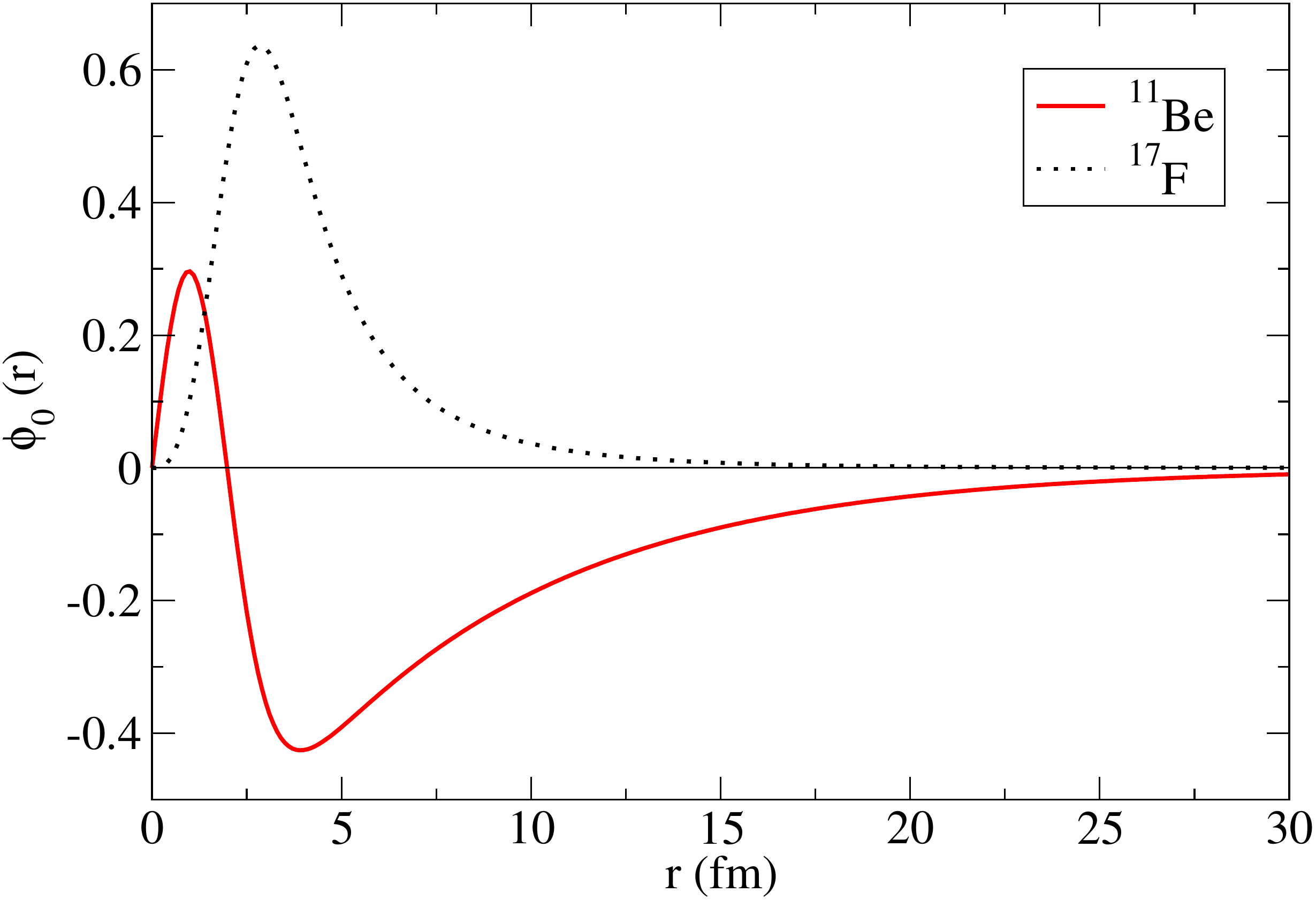}
\caption{Ground state wave function of the $^{11}$Be and $^{17}$F as a 
function of radial distance (r).}
\label{fig8a}
\end{center}
\end{figure}
\begin{figure}[htbp]
\begin{center}
\includegraphics[width=7.1cm]{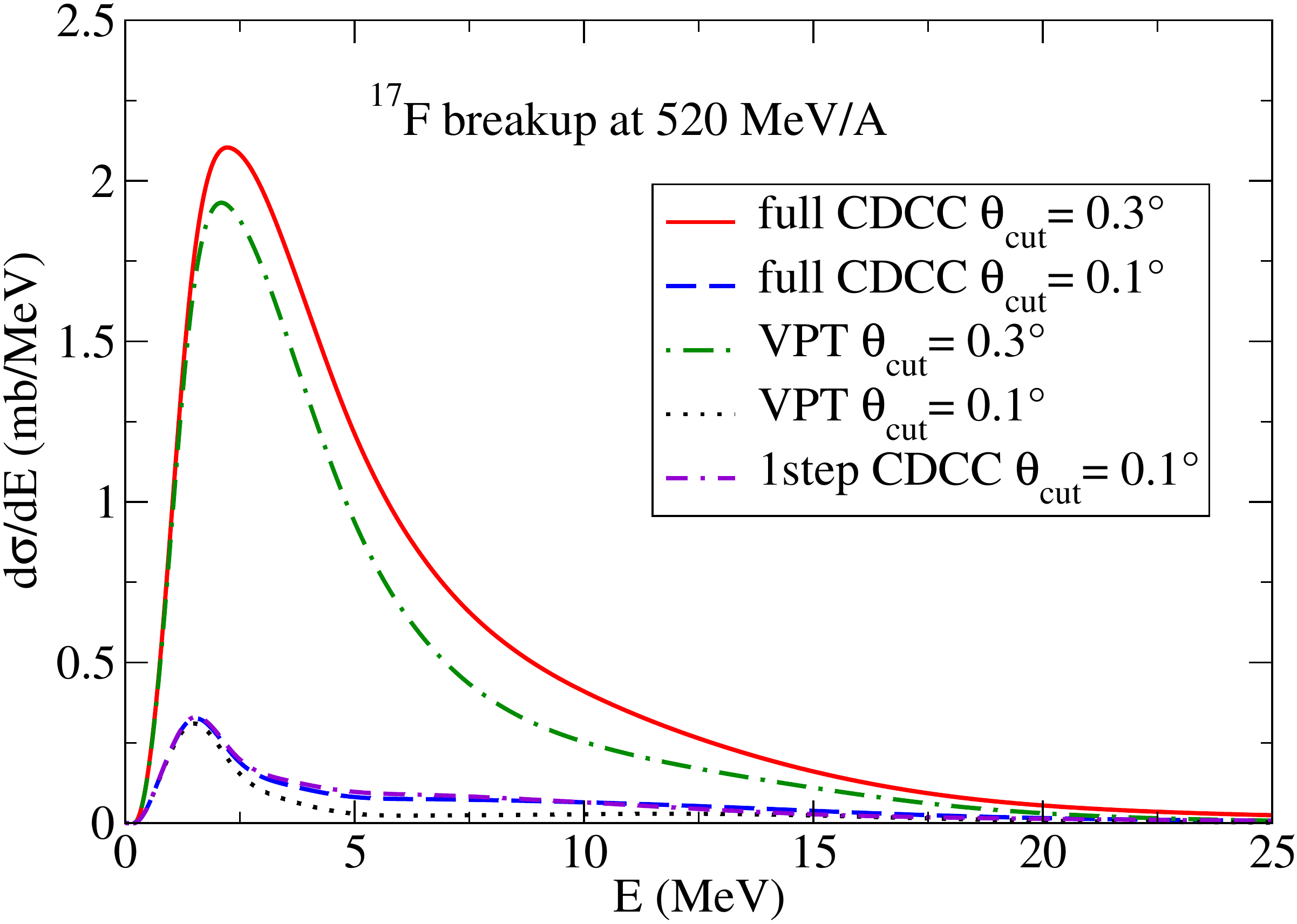}
\caption{Breakup cross section of $^{17}$F on $^{208}$Pb target at $520$\,MeV/A. 
The full CDCC (solid and dashed lines), VPT (dot-dashed and dotted lines), and 1step CDCC (dash-dotted line) results for different choices of the cutoff of 
the scattering angle, $\theta_{cut}$, are plotted.}
\label{fig8b}
\end{center}
\end{figure}

Let us discuss $^{17}$F breakup on $^{208}$Pb at $100$\,MeV/A shown in Fig.~\ref{fig8}~(a) and (b) for different settings.
In Fig.~\ref{fig8} (a), it can be clearly seen that the VPT agrees well with the full 
CDCC calculation at very small excitation energies ($E$ $\le 1.0$\,MeV), 
whereas a significant difference can be seen around the peak and 
in the tail region. The significant difference of the VPT with 
full CDCC calculation, clearly indicates the importance of missing contribution 
from the nuclear breakup and multistep processes.
It is shown in Fig.~\ref{fig8}~(a) that 1step CDCC also gives a notably larger cross section than full CDCC 
in the same region, which confirms the indication of VPT faliure.

To clarify the picture, the results corresponding to Coul, 1step Coul, and VPT 
calculations are shown in Fig.~\ref{fig8}~(b). 
The difference between Coul and 1step Coul confirms that the Coulomb driven breakup shows 
a significant multistep effect. 
Also, the difference between 1step Coul and VPT calculation reflects a considerable contribution from higher multipoles 
such as $E2$, which are missing in the VPT. 
These features can be attributed to the appearance of the additional Coulomb term in the $p$-T interaction 
compared with the $n$-T interaction. 
Similar to the breakup of C-$n$ projectiles, the significant multistep effect on nuclear breakup is found 
(shown in Fig.~\ref{fig8N}). 
Thus, the difference between 1step CDCC and full CDCC is due to the combined effect of both 
multistep Coulomb and nuclear breakup processes. 

The difference between the solid and dot-dashed lines in Fig.~\ref{fig8}~(a), 
shows that CNI effect is more prominent in the whole region as compared to the C-$n$ projectiles. 
Thus, in the breakup of $^{17}$F, along with various non-negligible higher-order processes 
found in the breakup of C-$n$ projectiles, the contribution from the higher multipoles and multistep effect 
on the Coulomb breakup, are found to play an important role. 

The same features can be seen at higher incident energies in 
Fig.~\ref{fig8} (c), (d), (e), and (f). 
From these findings, we remark that the assumptions adopted in the VPT are also not valid 
for a projectile with a C-$p$ structure.

From the C-$n$ projectiles to the C-$p$ projectiles, the major change is the appearance of the 
Coulomb term in the C-$p$ and $p$-T interaction. The role of the Coulomb term in the C-$p$ interaction 
is to constrain the $p$ closer to the C and as a consequence of this, their ground state wave functions have shorter 
tails with respect to C-$n$ cases (shown in Fig.~\ref{fig8a}). We also confirm this 
from the expectation value of the relative distance between the core and valence nucleon, which are found to be 
$6.97$ 
and $3.59$\,fm for $^{11}$Be and $^{17}$F, respectively.
Thus, C-$p$ breakup cross sections are smaller in magnitude compared to the C-$n$ breakup cross sections. 
However, for C-$p$ projectiles, Coulomb part of $p$-T interaction was found to play a role in the multistep effect.
The differences due to the multistep processes for $^{17}$F are about $37\%$, $14\%$, and $12\%$ at $100$, 
$250$, and $520$\,MeV/A, respectively. 
Unlike the C-$n$ cases, this is found to come from the combined effect of the 
energy dependence of the C-T and $p$-T nuclear potentials with the $p$-T Coulomb potential.
\begin{table}[htbp]
\caption{The integrated breakup cross sections in millibarn (mb) for 
$^{17}$F breakup on $^{208}$Pb target at $100$, $250$ and $520$\,MeV/A. 
The numerical integration is performed up to 
$25$\,MeV.}
\centering
\begin{tabular}{cccccc}
\hline\hline\\[-1.5ex]
Setting        &total          &$s$     &$p$       &$d$      &$f$   \\[1ex]
\hline\hline\\[-2.0ex]
&\multicolumn{5}{c}{\underline{$^{17}$F $+$ $^{208}$Pb at 100 MeV/A}}\\[1ex]
full CDCC&44.85&0.90&21.55&8.09&14.31\\
1step CDCC&61.45 ($+$37\%)&1.07&26.79&11.97&21.63 \\
VPT&57.30 ($+$27\%)&0.00&31.62&0.12&25.56\\
Coul&43.38 ($-$3\%)&0.64&23.88&3.23&15.63\\
1step Coul&62.27 ($+$41\%)&0.71&32.12&3.58&25.86\\
Nucl&16.57&0.63&3.55&6.96&5.43\\
1step Nucl&28.28&1.41&6.35&11.94&8.58\\
Coul+Nucl&59.95 ($+$34\%)&1.41&27.43&10.19&21.06\\
\hline\hline\\[-2.0ex]
\\ [-2.0ex]
&\multicolumn{5}{c}{\underline{$^{17}$F $+$ $^{208}$Pb at 250 MeV/A}}\\[1ex]
full CDCC&45.72&0.37&20.03&4.09&21.23 \\
1step CDCC& 52.23 ($+$14\%)&0.46&22.08&4.01& 25.68\\
VPT& 53.46 ($+$17\%)&0.00&24.31&0.08&29.07 \\
Coul& 46.49 ($+$2\%)&0.35&  21.68&2.07&22.39 \\
1step Coul& 56.18 ($+$23\%)&0.38&24.58&2.07&29.15 \\
Nucl&5.30&0.18&1.03&2.55&1.54\\
1step Nucl&6.78&0.30&1.45&3.03&2.00\\
Coul+Nucl& 51.79 ($+$13\%)&0.53&22.71&4.62&23.93 \\
\hline\hline\\[-2.0ex]
\\ [-2.0ex]
&\multicolumn{5}{c}{\underline{$^{17}$F $+$ $^{208}$Pb at 520 MeV/A}}\\[1ex]
full CDCC&51.34&0.48&18.56&6.10&26.20 \\
1step CDCC&57.20 ($+$12\%)&0.92&20.63&6.89&28.76\\
VPT&49.51 ($-$4\%)&0.00&19.72&0.08&29.71\\
Coul&45.51 ($-$11\%)&0.25&18.59&1.50&25.17\\
1step Coul&51.42 ($+$0.2\%)&0.26&19.92&1.47&29.77\\
Nucl&11.16&0.31&2.33&4.64&3.88\\
1step Nucl&16.34&0.75&3.92&6.07&5.60\\
Coul+Nucl&56.67 ($+$10\%)&0.56&20.92&6.15&29.04\\


\hline\hline
\end{tabular}
\label{T4}
\end{table}

To discuss the dependence of the breakup cross section on the scattering 
angle $\theta$ of the C-$p$ center of mass system, full CDCC, VPT, and 1step CDCC results for different 
choices of the $\theta_{cut}$ at $520$\,MeV/A is shown in Fig.~\ref{fig8b}. It can be clearly seen that with 
$\theta_{cut}=0.3\degree$, the results of full CDCC and VPT are still different around the peak and in the tail region. 
At $\theta_{cut}=0.1\degree$, the difference becomes negligible and the result of 1step CDCC also agrees well with the other two. 
This confirms the conclusion made in the previous section, that the choice of $\theta_{cut}$ depends on both the system under investigation and 
beam energy.

In Table~\ref{T4}, we give the total and partial waves ($s$, $p$, $d$, and $f$) integrated breakup cross sections 
calculated with different settings.
It can be clearly seen from the Table~\ref{T4}, for Coul$+$Nucl, CNI is significantly constructive in all the partial waves individually 
as well as in total, at each incident energy. 

We confirmed our findings for other C-$p$ projectile, $^{8}$B with different ground state 
configuration \textit{i.e.}, $p$-wave. Qualitatively, for the role of the nuclear breakup contribution, CNI effect, and 
multistep processes, we have observed the same features as for $^{17}$F. Quantitatively, the magnitude of breakup 
cross sections are larger, which is due to the lower binding energy of the $^{8}$B as compared to the $^{17}$F. 

In summary, with respect to C-$n$ cases, in the breakup of C-$p$ projectiles, the source of the multistep effects 
is the combined effect of the Coulomb and nuclear interaction. Also, in the Coulomb driven breakup for C-$p$ projectiles, 
the higher multipoles such as $E2$ plays a significant role. CNI effects are found to be more prominent in the breakup of C-$p$ projectiles.

\section{Summary and Conclusions}\label{Sum}
We reported the systematic investigation on the role of various higher-order processes, which are ignored in the VPT, 
such as the contribution of nuclear breakup, higher multipoles in the Coulomb breakup, CNI effect, 
and multistep processes due to strong continuum-continuum couplings.
We conducted this study at the intermediate ($100$ and $250$\,MeV/A) and higher ($520$\,MeV/A) incident energies, 
in the breakup of different two-body projectiles, having core-neutron ($^{11}$Be) and core-proton ($^{17}$F) 
structure on a heavy target ($^{208}$Pb), using the E-CDCC method.

Our results showed that, for all the projectiles, the nuclear breakup contribution is important and 
it plays a significant role in the multistep processes. 
The multistep effect on Coulomb breakup is found to be negligible for C-$n$ projectiles. 
For C-$p$ projectiles, multistep processes appears from the combined effect of 
the Coulomb and nuclear breakup parts, and also for Coulomb breakup, higher multipoles, greater than $E1$, 
found to play significant role. CNI effect, appearing from incoherent sum of Coulomb and nuclear breakup contribution, is found to be non-negligible and  
is more prominent for C-$p$ projectiles than the C-$n$ projectiles. 

Our detailed investigation shows that the nuclear breakup component, higher multipoles in the Coulomb breakup, the CNI effect, 
and the multistep breakup processes are all found to be non-negligible. 
From these findings, we conclude that the assumptions adopted in the VPT are not valid for accurate 
description of breakup cross sections at intermediate and higher incident energies . Another important conclusion of our study is, quantitatively, 
the multistep effects due to the nuclear breakup are found to depend on the incident energy via 
the energy dependence of the C-T and N-T nuclear potentials.

Additionally, the choice of some cut on the scattering angle $\theta$, assuming that the nuclear breakup 
contribution can be ignored below 
that cutoff angle $\theta_{\rm cut}$ is also investigated. Our results showed that 
the choice of $\theta_{cut}$ depends on the system under investigation and 
beam energy.

We strongly expect that our study will provide support to the various ongoing experimental studies on the 
breakup of light and medium-mass nuclei, at various experimental 
facilities. As a future perspective, it is very interesting to extend our study to the breakup of 
three-body projectiles such as $^6$He, $^{11}$Li, $^9$C, $^{22}$C, $^{29}$F which we intend to report elsewhere in near future.

\section*{Acknowledgments}
The authors would like to thank T. Fukui and NTT Phuc, for fruitful comments and discussions.
The computation was carried out using the computer facilities at the
Research Center for Nuclear Physics, Osaka University.
This work was supported in part by Grants-in-Aid of the Japan Society for the Promotion of Science (Grant No. JP18K03650).
\section*{Corrections/Changes w.r.t version1}
\subsection*{Typos in version1}
\begin{itemize}
\item In supplemental material for Fig.~S16, inside the panel (b), \textquotedblleft$100$\,MeV/A $\rightarrow$ $250$\,MeV/A\textquotedblright.
\item In supplemental material for Fig.~S18, inside the panel (a), \textquotedblleft$520$\,MeV/A $\rightarrow$ $100$\,MeV/A\textquotedblright.
\end{itemize}
\subsection*{Changes in version2}
\begin{itemize}
\item  This version includes discussion on two cases only, i.e., $^{11}$Be and $^{17}$F. 
\vspace*{-0.1cm}
\item For $^{31}$Ne and $^8$B results, refer to  arXiv:2005.05605v1.
\vspace*{-0.1cm}
\item In this version, for $^{11}$Be, the calculations include the spin of valence 
 neutron and comparison is made with experimental data for $520$\,MeV/A case.
\end{itemize}


\begin{thebibliography}{xxx}
\bibitem{Bennet00} R. Bennett \textit{et al.}, NuPECC Report on and Radioactive Nuclear Beam Facilities, April (2000).
\bibitem{Tanihata85} I. Tanihata \textit{et al.}, Phys. Rev. Lett. \textbf{55}, 2676 (1985).
\bibitem{Tanh85} I. Tanihata \textit{et al.}, Phys. Lett. B {\bf160}, 380 (1985).
\bibitem{Otsuka01} T. Otsuka \textit{et al.}, Phys. Rev. Lett. \textbf{87}, 082502 (2001).
\bibitem{Kobayashi88} T. Kobayashi \textit{et al.}, Phys. Rev. Lett. \textbf{60}, 2599 (1988).
\bibitem{Fukuda91} M. Fukuda \textit{et al.}, Phys. Lett. B {\bf268}, 339 (1991).
\bibitem{Palit03} R. Palit \textit{et al.}, Phys. Rev. C \textbf{68}, 034318 (2003).
\bibitem{Fukuda04} N. Fukuda \textit{et al.}, Phys. Rev. C \textbf{70}, 054606 (2004).
\bibitem{Warner95} R.~E. Warner \textit{et al.}, Phys. Rev. C \textbf{52}, R1166(R) (1995).
\bibitem{Negoita96} F. Negoita \textit{et al.}, Phys. Rev. C \textbf{54}, 1787 (1996).
\bibitem{Bazin98} D. Bazin \textit{et al.}, Phys. Rev. C \textbf{57}, 2156 (1998).
\bibitem{Gulmaraeas00} V. Guimar\~{a}es \textit{et al.}, Phys. Rev. C \textbf{61}, 064609 (2000).
\bibitem{TOG16} Y. Togano {\it et al.}, Phys. Lett. B {\bf761}, 412  (2016).
\bibitem{Nakamura2009} T. Nakamura \textit{et al.}, Phys. Rev. Lett. \textbf{103}, 262501 (2009).
\bibitem{Gaudefroy2012} L. Gaudefroy \textit{et al.}, Phys. Rev. Lett. \textbf{109}, 202503 (2012).
\bibitem{Kobayashi2014} N. Kobayashi \textit{et al.}, Phys. Rev. Lett. \textbf{112}, 242501 (2014).
\bibitem{Bertulani10} C.~A. Bertulani and A. Gade, Phys. Rep. \textbf{485}, 195 (2010).
\bibitem{Langanke13} K. Langanke and H. Schatz, Phys. Scr. T \textbf{152}, 014011 (2013).
\bibitem{Bertulani16} C.~A. Bertulani and T. Kajino, Prog. Part. Nucl. Phys. \textbf{89}, 56 (2016).
\bibitem{Yahiro12} M. Yahiro K. Ogata, Takuma Matsumoto and K. Minomo, Prog. Theor. Exp. Phys. \textbf{01A206} (2012).
\bibitem{Chatterjee18} R. Chatterjee and R. Shayam, Prog. Part. Nucl. Phys. \textbf{103}, 67 (2018).
\bibitem{Bonaccorso18} A. Bonaccorso, Prog. Part. Nucl. Phys. \textbf{101}, 1 (2018).
\bibitem{Bertulani88} C.~A. Bertulani and G. Baur, Phys. Rep. \textbf{163}, 299 (1988).
\bibitem{Baur96} G. Baur and H. Rebel, Annu. Rev. Nucl. Part. Sci. \textbf{46}, 321 (1996).
\bibitem{Baur03} G. Baur, K. Hencken, and D. Trautmann, Prog. Part. Nucl. Phys. \textbf{51}, 487 (2003).
\bibitem{Tanihata95} I. Tanihata, Prog. Part. Nucl. Phys. \textbf{35}, 505 (1995) and references cited therein.
\bibitem{Hus06} M.~S. Hussein, R. Lichtenth\"aler, F.~M. Nunes, and I.~J. Thompson, Phys. Lett. {\bf B640}, 91 (2006).
\bibitem{Thompson09} I.~J. Thompson and F.~M. Nunes, \textit{Nuclear Reactions for Astrophysics} (Cambridge University Press, New York, 2009).
\bibitem{Chatterjee02} R. Chatterjee and R. Shyam, Phys. Rev. C \textbf{66}, 061601(R) (2002).
\bibitem{Chatterjee03} R. Chatterjee, Phys. Rev. C \textbf{68}, 044604 (2003).
\bibitem{Chatterjee07} R. Chatterjee, Phys. Rev. C \textbf{75}, 064604 (2007).
\bibitem{Tarutina04} T. Tarutina and M.~S. Hussein, Phys. Rev. C \textbf{70}, 034603 (2004).
\bibitem{Mukeru15} B. Mukeru, M. L. Lekala, and A. S. Denikin, J. Phys. G: Nucl. Part. Phys. \textbf{42}, 015109 (2015).
\bibitem{Dasso98} C.~H. Dasso, S.~M. Lenzi, and  A. Vitturi, Nucl. Phys. A \textbf{639}, 635 (1998).
\bibitem{Nunes98} F.~M. Nunes and I.~J. Thompson, Phys. Rev. C \textbf{57}, R2818(R) (1998).
\bibitem{Nunes99} F.~M. Nunes and I.~J. Thompson, Phys. Rev. C \textbf{59}, 2652 (1999).
\bibitem{Rangel16} J. Rangel, J. Lubian, L.~F. Canto, and P.~R.~S. Gomes, Phys. Rev. C \textbf{93}, 054610 (2016). 
\bibitem{Kumar11} R. Kumar and A. Bonaccorso, Phys. Rev. C \textbf{84}, 014613 (2011). 
\bibitem{Kucuk12} Y. Kucuk and A.~M. Moro, Phys. Rev. C \textbf{86}, 034601 (2012). 
\bibitem{Paes12} B. Paes, J. Lubian, P.~R.~S Gomes, and V. Guimar\~{a}es, Nucl. Phys. A \textbf{890-891}, 1 (2012).
\bibitem{Mukeru18} B. Mukeru, J. Phys.  G: Nucl. Part. Phys. \textbf{45}, 065201 (2018). 
\bibitem{Ogata09} K. Ogata and C.~A. Bertulani, Prog. Theor. Phys. \textbf{121}, 1399 (2009).
\bibitem{Ogata10} K. Ogata and C.~A. Bertulani, Prog. Theor. Phys. \textbf{123}, 701 (2010).
\bibitem{Laura19} L. Moschini and P. Capel, Phys. Lett. B {\bf790}, 367 (2019).
\bibitem{Ogata03} K. Ogata, M. Yahiro, Y. Iseri, T. Matsumoto, and M. Kamimura, Phys. Rev. C \textbf{68}, 064609 (2003).
\bibitem{Ogata06} K. Ogata, S. Hashimoto, Y. Iseri, M. Kamimura, and M. Yahiro, Phys. Rev. C \textbf{73}, 024605 (2006).
\bibitem{Nakamura17} T. Nakamura \textit{et al.}, Prog. Part. Nucl. Phys. \textbf{97}, 53 (2017)and references cited therein.
\bibitem{Rahman17} A. Rahaman \textit{et al.}, J. Phys. G: Nucl. Part. Phys. \textbf{44}, 045101 (2017).
\bibitem{Cook20} K. Cook \textit{et al.}, Phys. Rev. Lett. \textbf{124}, 212503 (2020).
\bibitem{Desco17} P. Descouvemont, L.~F. Canto, and M.~S. Hussein, Phys. Rev. C \textbf{95}, 014604 (2017).
\bibitem{Capel04} P. Capel, G. Goldstein, and D. Baye, Phys. Rev. C \textbf{70}, 064605 (2004).
\bibitem{Spa00} J.~M. Sparenberg, D. Baye, and B. Imanishi, Phys. Rev. C \textbf{61}, 054610 (2000).
\bibitem{Amo00} K. Amos, P.~J. Dortmans, H.~V. von Geramb, S. Karataglidis, and J. Raynal, Adv. Nucl. Phys. {\bf 25}, 275 (2000). 
\bibitem{DG80} J. Decharge and D. Gogny, Phys. Rev. C \textbf{21}, 1568 (1980).
\bibitem{Ber91} J.~F. Berger, M. Girod, and D. Gogny, Comp. Phys. Comm. \textbf{63}, 1365 (1991).
\bibitem{Min12} K. Minomo, T. Sumi, M. Kimura, K. Ogata, Y.~R. Shimizu, and M. Yahiro, Phys. Rev. Lett. \textbf{108}, 052503 (2012).
\bibitem{Sum12} T. Sumi, K. Minomo, S. Tagami, M. Kimura, T. Matsumoto, K. Ogata, Y.~R. Shimizu, and M. Yahiro, Phys. Rev. C. {\bf 85}, 064613 (2012).

\end{thebibliography}
\end{document}